\documentclass[amsmath, amssymb, aip, showkeys, twocolumn,reprint]{revtex4-2}
\usepackage[T1, T2A]{fontenc}
\usepackage[english]{babel}

\usepackage{caption}
\usepackage{subcaption}
\usepackage{graphicx}
\usepackage{placeins}
\usepackage{xcolor}
\usepackage{comment}
\usepackage{listings}
\usepackage{natbib}
\usepackage{appendix}

\usepackage{graphicx}
\graphicspath{{images/}}
\DeclareGraphicsExtensions{.pdf,.png,.jpg}
\usepackage{physics}
\usepackage{amssymb,amsfonts,amsmath,amsthm, bm}
\usepackage[unicode=true]{hyperref}

\newcommand{\Fr}{\operatorname{Fr}}

\newcommand{\Grad}{\operatorname{grad}}
\newcommand{\Curl}{\operatorname{curl}}

\newcommand{\dV}{\delta\bm{V}}

\newcommand{\dtVx}{\delta\tilde{V}_x}
\newcommand{\dVy}{\delta{V}_y}

\newcommand{\dVz}{\delta{V}_{\!z\,}}
\newcommand{\dtVz}{\delta\tilde{V}_z}
\newcommand{\dO}{\delta\bm{\Omega}}
\newcommand{\dOx}{\delta{\Omega}_x}
\newcommand{\dtOx}{\delta\tilde{\Omega}_x}
\newcommand{\dPsi}{\delta{\Psi}}
\newcommand{\dtPsi}{\delta\tilde{\Psi}}
\newcommand{\dpressure}{\delta{p}^{\varpi}}
\newcommand{\zero}{{\scriptscriptstyle{(0)}}}
\newcommand{\Ozero}{\Omega^{\scriptscriptstyle{(0)}}}
\newcommand{\hzero}{h^{\scriptscriptstyle{(0)}}}

\newcommand{\nutu}{\nu }

\begin{document}
\sloppy
\title{Role of wave scattering in instability-induced Langmuir circulation\\ {\normalsize {\normalfont Published in: Physics of Fluids} \textbf{36}{\normalfont, 034119 (2024)}}}
\author{Sergey\,S.~Vergeles$^{1,2}$}
\email{ssver@itp.ac.ru}
\author{Ivan\,A.~Vointsev$^{1,2}$}
\email[Author to whom correspondence should be addressed:\ ]{ivointsev@itp.ac.ru}
\affiliation{$^{1}$L.\,D.~Landau Institute for Theoretical Physics RAS, Prosp. Akademika Semenova 1A, Chernogolovka, Moscow region, 142432, Russian Federation \\
$^{2}$National Research University Higher School of Economics, Faculty of Physics, Myasnitskaya 20, 101000 Moscow, Russia}
\date{\today}

\begin{abstract}
We consider a classical problem about dynamic instability that leads to the Langmuir circulation. The problem statement assumes that there is initially a wind-driven shear flow and a plane surface wave propagating in the direction of the flow. The unstable mode is a superposition of i) shear flow and ii) surface waves both modulated in the horizontal spanwise direction and iii) circulation that is made up with vortices forming near-surface rolls whose axis are coaligned along the shear flow
streamlines and whose transverse size corresponds to the modulation period. Usually, the Langmuir circulation is understood as the vortical part of the mode slowly varying in time, which is the combination of the first and the last flows. The novelty of our approach is that we, firstly, take into account the scattering of the initial surface wave on the slow current. Second, we find the interference of the scattered and the initial waves generating a Stokes drift modulated in the same direction. Third, we establish the subsequent affect of the circulation by the vortex force created by the nonlinear interaction of the initial shear flow and the modulated part of the Stokes drift. S. Leibovich \& A.D.D. Craik previously showed that the third part of the mechanism could maintain the Langmuir circulation. We calculate the growth rate which is approximately twice smaller than that obtained by A.D.D.\,Craik. The vertical structure of the circulation in the mode consists of two vortices, that corresponds to the next mode in Craik's model. Considering the wave scattering, we describe the fast-wave motion as a potential flow with relatively weak vortical correction. Application of the technique can be expanded on other flows where fast oscillating surface waves coexist with a slow current.
\end{abstract}

\maketitle

\section{Introduction}

Langmuir circulation (LC) is an important mechanism which is responsible for enhancement of turbulent mixing in the ocean surface boundary layer, see e.g. Ref.~\onlinecite{hamlington2014langmuir} and reviews~\onlinecite{thorpe2004langmuir,teixeira2019}. The circulation arises on the background flow generated by a plane surface wave propagating along a vertical shear flow, the latter is typically driven by a coaligned air wind. It includes vortical near-surface rolls whose axes are oriented in the wave-wind direction and a codirectional shear flow modulated in the horizontal spanwise direction. Although the most interesting question is the description of established LC \cite{WELLER1988711, chang2019small}, the process of the emergence and the growth of the circulation are also of certain interest \cite{craik1977generation, leibovich2009langmuir}, as they provide a tool for the analysis of experimental \cite{faller1978laboratory,smith1992observed} and numerical \cite{sullivan_mcwilliams_2019} data. This emergence could be described by an instability mechanism. If one drops out wave scattering on the circulation, the following positive feedback leads to the instability. Consider a small spanwise modulation of the shear flow. This modified flow possesses a vertical vorticity in contrast to the initial shear flow. The wave affects the modulated current, the impact is described by a vortex force \cite{craik_leibovich_1976}, which is the cross product of the Stokes drift produced by the wave motion and the vertical vorticity of the slow current. Thus, the vertical vorticity and the Stokes drift produce a spanwise vortex force that rolls the flow up. The latter interacts with the initial shear flow via the Lamb term in the Navier-Stokes equation, which drives the spanwise modulation of the shear flow, so the feedback loop is closed.

Craik’s analysis \cite{craik1977generation} did not consider the wave scattering on the modulated flow, although it exists and leads to the spanwise modulated wave according to the further analytical investigations \cite{craik_1982}. This modulation was indeed observed in the large variety of experimental and numerical works. For example, large-eddy simulations of interacting turbulent shear flow and surface waves \cite{kawamura2000numerical} showed rather noticeable variation of wave field in the crosswind direction that cannot be omitted when describing oceanic boundary layer. In Ref.~\onlinecite{veron2001experiments}, the strong LC-like modulation of the wave field was proved during laboratory experiments. The results of recent numerical calculations also indicate that the wave scattering by the background flow and the subsequent wave interference cannot be neglected \cite{fujiwara2020mutual}. Some analytical study of coupling between a surface wave and the circulation can be found in Ref.~\onlinecite{suzuki2019physical}. A general conclusion is that the influence of established LC on the surface motion is an important factor in the evolution of the ocean surface boundary layer. The effect of the interference of two waves of equal magnitudes and opposite signs, propagating at small angles with respect to the shear flow was considered separately in two analytical approaches\cite{craik_leibovich_1976,leibovich1977evolution} and was shown to cause linear LC growth, if this LC is absent or weak, and to maintain the circulation when the growth is saturated due to the turbulent viscosity. In this paper we show that the scattering leads to the weakening of the positive feedback so the instability develops slower than it is predicted in Ref.~\onlinecite{craik1977generation}.

In Ref.~\onlinecite{craik_1982}, the LC-induced wave modulation was considered and the instability problem was examined in the limit of short spanwise modulation space scale compared to the wavelength. In the limit, no wave scattering occurs since the wavenumber of the modulation much exceeds the wavenumber of the initial wave, and the scattered wave would have almost the same frequency as the initial one because the time scale associated with the slow flow is much greater compared to the wave period. One can say, that the slow current is a kind of static and periodic in space impact on the wave, so an analogue for the wave scattering in our problem is surface wave scattering on periodically rippled bed for finite-depth water \cite{couston2017shore}. The wave scattering is possible only if the period is comparable to or larger than the wavelength. Here we consider this limit of short wavelength compared to the modulation period, so the initial wave may scatter on the circulation and the scattered wave should be taken into account. The stream function used in Ref.~\onlinecite{craik_1982} is convenient when there is a single wave vector characterising the wave motion. As we essentially consider a three-dimensional wave flow consisting of waves spreading in different directions, we use a potential approximation for the flow. When describing the wave interaction with the slow current, the relatively weak vortical correction to the wave flow should be found \cite{craik_leibovich_1976}. This technics allows us to consistently take into account the wave scattering on a vortex flow. The only requirement for the applicability of the general mathematical tools is that the wave frequency has to be much greater than the velocity gradient and the inverse characteristic time scale of the vortex flow. The time scale separation allows one to consider at the leading order, the wave motion and the vortical flow as two independent subsystems that weakly interact through the Navier-Stokes equation nonlinearities. In the Langmuir instability problem, we show that the spanwise modulation of the wave is essential whatever the shear flow strength may be. In particular, we obtain the unstable mode growth rate smaller than that found in Ref.~\onlinecite{craik1977generation}. The vertical structure of the circulation in the mode consists of two vortices with opposite signs. At a qualitative level, the wave scattering stops the growth of the main Craik's mode, so that the fastest growing mode becomes the next one, which has just one vortex located above the other. In our model, the mode is deformed by the influence of the scattered wave. According to the general requirement, the established analytical results are applicable only if the shear rate is much less than the wave frequency. The additional assumption is the smallness of the Langmuir number, that allows one to analyze the inviscid problem.

The general mathematical scheme is developed in Section~\ref{sec:general}. The derivation of the vortex force is reproduced in Subsection~\ref{subsec:vortex-force} and the wave scattering process is considered in Subsection~\ref{subsec:wave-scattering}. Using this technique, the Langmuir instability mechanism is examined in Section~\ref{sec:Langmuir-instability}. The unperturbed and perturbation flows are respectively defined in Subsections~\ref{subsec:unperturbed-flow},\,\ref{subsec:perturbation}. The equations governing the linearized development of the perturbation are derived in Subsection~\ref{subsec:linearized-equations}. Then, we consider the small-Langmuir number limit, $\mathrm{La}\ll1$ and present a complete analytical solution of the problem in the linear-shear case in Subsection~\ref{subsec:LC-solution}. Some calculations are relegated to into Appendix.

\section{General description of interaction between waves and slow current}
\label{sec:general}

We consider a flow of an incompressible fluid with a free surface. The surface coincides with the plane $z=0$ if the fluid is at rest, where $Oxyz$ is a Cartesian coordinate system, and it is described by the function $z = \zeta (x, y, t)$ if the flow is excited. The fluid motion is described by the Navier-Stokes equation that can be written in the Lamb form\cite{lamb1916hydrodynamics}
\begin{equation}\label{lamb}
    \partial_t {\bm v} = [\bm{v}, \,\bm{\varpi}]
    -
    \Grad\left(p+ \frac{v^2}{2}\right)
    +
    \nu\Delta\bm{v} + \bm{g},
\end{equation}
where ${\bm v}$ is the velocity field, $p$ is the fluid pressure, ${\bm \varpi}=\mathop{\mathrm{curl}}{\bm v}$ is the vorticity, the mass density is equal to one and \textquoteleft$[\star,\star]$\textquoteright\ means the cross product of two vectors. The equation should be supplemented with boundary conditions that are, first the following kinematic boundary condition
\begin{equation}\label{kinematic}
    \partial_t \zeta
    =
    v_z - v_{\alpha}\partial_{\alpha}\zeta,
\end{equation}
where indices $\alpha,\beta,\ldots$ stand for the coordinates $\{x,y\}$ and, second the following dynamic boundary conditions
\begin{align}
    \label{dynamic_pressure}
    \big(p - 2\nu n_{i}n_{k}\partial_{i}v_{k}\big)\big|_{z = \zeta} ={}& 0, \\[5pt]
    \label{dynamic_tangent}
    \delta^{\scriptscriptstyle \perp}_{l i}\,(\partial_{i}v_{k} + \partial_{k}v_{i})n_{k}\big|_{z = \zeta} ={}& 0,
\end{align}
where $\delta^{\scriptscriptstyle \perp}_{l i} = \delta_{li} - n_ln_i$ is the projector onto the plane tangent to the free surface, ${\bm n}$ is the normal unit vector on the surface and the indices $i,j,k,\ldots$ stands for coordinates $\{x,y,z\}$. Hereinafter, summation is assumed over the repeating indices. The liquid is assumed to be unlimitedly deep.

The velocity field ${\bm v}$ is a sum of surface gravity wave flow ${\bm u}$ and vortical current ${\bm V}$, i.e. $\bm{v} = \bm{u} + \bm{V}$. The characteristic time scale $T$ of the vortical current, $T\sim |{\bm V}|/|\partial_t{\bm V}|$ or $T\sim 1/|\mathop{\mathrm{grad}}{\bm V}|$, is much larger than the inverse gravity wave frequency $\omega$, $\varepsilon = 1/\omega T \ll 1$, so we call ${\bm u}$ high-frequency component of the whole flow and the vortical current ${\bm V}$ is the slow part of the flow. Next, the slowness of the current assumes that it does not lead to variation in the free surface form \cite{holm1996ideal}, so $\zeta$ contains only the high-frequency component. The criterion is the Froude number of the slow current should be small, $\Fr \sim |{\bm V}| \cdot|\mathop{\mathrm{grad}}{\bm V}|/g \ll 1$, where $g$ is the gravitational acceleration. Note that the spatial scale of the current $L\sim|{\bm V}|/|\mathop{|\mathrm{grad}}{\bm V}|$ is not assumed to be necessarily large in relation to the wavelength, but the smallness of parameter $\varepsilon$ is an essential condition for our analytical approach to be applicable.

In the current and viscosity absences, the non-breaking wave flow u is irrotational,
so it can be described in term of a flow potential $\phi$. The interaction between the wave and the vortical current leads the fast component of the flow oscillating with the wave frequency to cease to be
purely irrotational. There also exists an analytical description of the phenomenon based
on the wave flow representation through the stream function instead of the potential $\phi$,
which is good for two-dimensional problems \cite{stewart1974hf}. When the wave motion is substantially three-dimensional, the technique however becomes cumbersome since it is impossible to describe the wave flow with an only scalar stream function, so one should introduce more functions \cite{craik_1982}. Our aim is to describe the dynamics of the wave motion itself including its interaction with the current. The results from the interaction at large distances are scattered waves that can be reasonably described in terms of the potential $\phi$. There also exists a local response to the interaction, which is relatively small as $\varepsilon$ and can be described in terms of oscillating vorticity $\bm{\varpi}^{u} = \Curl{\bm{u}}$ and pressure part $p^{\varpi}$ (see below). Thus in our approach, we keep the potential $\phi$ which determines the wave motion itself and the vortical correction $\bm{u}^{\varpi}$ so that
\begin{equation}\label{u-division}
    \bm{u} = \bm{u}^{\phi} + \bm{u}^{\varpi},
    \quad \bm{u}^{\phi} = \Grad\phi,
    \qquad
    u^{\varpi} \sim\varepsilon\cdot u^\phi.
\end{equation}
For the separation (\ref{u-division}) into the irrotational and the vortical parts may be effective, we assume that $\bm{u}^{\varpi}$ is not related to the surface dynamics in the linear approximation
\begin{equation}\label{upsi-cindition}
    u^{\varpi}_{z}|_{z = 0} = 0,
\end{equation}
since the surface shape dynamics is closely related to the potential wave motion.

\subsection{Influence of wave motion on slow current}
\label{subsec:vortex-force}

The impact of the waves on the current is established in Refs.~\onlinecite{craik_leibovich_1976,leibovich1977evolution} and is called the vortex force. On the way of its derivation, one should find the high-frequency part ${\boldsymbol \varpi}^u$ of the vorticity, which is also used when finding the scattered waves as well. So here, we briefly repeat the derivation of the vortex force $\bm{f}^V$.

The projection of the Navier-Stokes equation (\ref{lamb}) onto the slow motion is
\begin{equation}\label{V-slow-NS}
    \partial_t\bm{V} + (\bm{V}\, \nabla)\bm{V}
    =
    -
    \Grad\left\langle p+ \frac{ u^2}{2}\right\rangle
    + \nutu \Delta{\bm V} +
    \langle [\bm{u}^\phi, \, \bm{\varpi}^u]\rangle
    ,
\end{equation}
where angle brackets $\langle \ldots\rangle$ stand for the time averaging over the fast wave oscillations. The last term in (\ref{V-slow-NS}) stems from the high-frequency part of the flow, which is purely potential in the main approximation, so ${\bm u}$ has been replaced by ${\bm u}^\phi$. To find the high-frequency part ${\boldsymbol \varpi}^u$ of the vorticity, we linearize the vorticity equation
\begin{equation}\label{vorticity-equation}
    \partial_t{\bm \varpi}
    =
    \mathop{\mathrm{curl}}[{\bm v},{\bm \varpi}]
    +
    \nu \Delta {\bm \varpi}
\end{equation}
with respect to the wave motion, which leads to
\begin{equation}\label{vorticity-equation-lin}
    \big(\partial_t + (\bm{V}\,\nabla)\big)\bm{\varpi}^{u} -  (\bm{\varpi}^{u}\nabla)\bm{V}
    - \nu\Delta\bm{\varpi}^{u}
    = \Curl\left[\bm{u}^{\phi},\, \bm{\Omega}\right],
\end{equation}
where the vorticity of the slow flow ${\boldsymbol \Omega}=\operatorname{curl}{\bm V}$, the full vorticity is therefore  ${\bm \varpi} = {\bm \Omega} + {\bm \varpi}^u$. The absolute value of the second term in the left-hand side of equation~\eqref{vorticity-equation-lin} can be estimated as $\left|(\bm{\varpi}^{u}\nabla)\bm{V} \right| \sim\Omega\varpi^{u}$ so the term should be neglected compared to $\partial_t \bm{\varpi}^{u}$ due to $|\partial_t \bm{\varpi}^{u}|\sim \omega \varpi^{u}$. The right-hand side serves as a driving force for ${\bm \varpi}^{u}$. We assume that the spatial scale $\sim\mathop{\mathrm{min}}(L.k^{-1})$ of the right-hand side of the equation is much greater than the thickness $\sqrt{\nu/\omega}$ of the viscous sublayer. Thus, the influence of the viscosity is negligible and the equation becomes local in space. Its solution is
\begin{equation}\label{vorticity-wave}
    \bm{\varpi}^{u} =
    \operatorname{curl}\left[\bm{s},\, {\bm \Omega}\right],
    \quad \bm{s}(t,{\bm r})=\int\limits^t dt' \, \bm{u}^{\phi}(t', \,\bm{r}(t')),
\end{equation}
where $\dot {\bm r}(t) = {\bm V}(t,{\bm r})$ is the Lagrangian trajectory produced by the slow flow, ${\bm r}(t) \equiv {\bm r}$ and ${\bm s}$ is a particle displacement during the wave oscillations. Note that all time averagings $\langle \ldots\rangle$ in (\ref{V-slow-NS}) and below in (\ref{vortex-force_new},\ref{Stokes}) should be implemented along the Lagrangian trajectories ${\bm r}(t)$. Using~\eqref{vorticity-wave}, one can rewrite the last term in~\eqref{V-slow-NS} into a more convenient form
\begin{equation}\label{vortex-force_new}
    \langle [\bm{u}^\phi, \, \bm{\varpi}^u]\rangle
    = \bm{f}^V + \frac{1}{2}\Grad (\bm \Omega \cdot \bm{\mathcal A}),
    \qquad
    \bm{f}^V = [\bm{U}^s, \, \bm{\Omega}],
\end{equation}
where $\bm{\mathcal A} = \langle [{\bm u}^\phi, {\bm s}]\rangle /2$ and the Stokes drift
\begin{equation}\label{Stokes}
    \bm{U}^s = \Curl\bm{\mathcal A} = \langle \Curl \, [{\bm u}^\phi, {\bm s}] \rangle/2
\end{equation}
is determined by the potential part of the flow associated with the wave motion. The term inside the gradient in (\ref{vortex-force_new}) should be included in the effective pressure in (\ref{V-slow-NS}).

Finally, the equation (\ref{V-slow-NS}) takes the form
\begin{equation}\label{V-equation}
   \partial_t\bm{V} + (\bm{V}\, \nabla)\bm{V}
    =
    -
    \Grad  \overline{P}
    +
    \nutu  \Delta {\bm V}
    +
    \bm{f}^V,
\end{equation}
where effective time-averaged pressure $\overline P$ contains terms inside gradients from (\ref{V-slow-NS}) and (\ref{vortex-force_new}). Concerning the boundary conditions, we neglect the virtual wave stress produced by the wave viscous damping \cite{longuet1953mass,nicolas2003three,filatov2016nonlinear}, which is a reasonable approximation if the vortex flow is strong enough so that the velocity gradient far exceeds the viscous damping rate of the wave, $|\nabla {\bm V}|\gg \nu k^2$. Then the boundary conditions for the slow flow are a stress-free rigid surface,
\begin{equation}\label{slow-flow-BC}
    \partial_z V_{\alpha}\big|_{z=0} = 0,
    \quad
    V_z\big|_{z=0}=0.
\end{equation}
Note that in the limit of ideal fluid, only the last condition in (\ref{slow-flow-BC}) should be kept.

\subsection{Wave scattering process}
\label{subsec:wave-scattering}

The wave flow dynamics is determined by the Navier-Stokes equation linearized with respect to the wave amplitude
\begin{equation}\label{u-fast-NS}
    \partial_t\bm{u}
    +
    \left(\bm{u}\nabla\right)\bm{V}
    +
    \left(\bm{V}\nabla\right)\bm{u}
    =
    -\nabla p^u+\bm{g},
\end{equation}
where $p^{u}$ is the high-frequency part of the pressure (for convenience, we included hydrostatic part of the pressure in $p^u$, which does no affect the slow current ${\bm V}$). We omitted the viscous term in (\ref{u-fast-NS}), so we deal with the Euler equation, due to the assumption that the viscosity which alters the flow in the narrow viscous sublayer beneath the surface is not essential in the wave dynamics. Our goal is to describe the dynamics of the potential part of the wave flow (\ref{u-division}) taking into account the scattering process. Due to the incompressibility condition, the potential still satisfies the Laplace equation
\begin{equation}\label{Laplace}
    \Delta\phi = 0.
\end{equation}

\begin{figure}[t]
    \centering
    \includegraphics[width = 1.0\linewidth]{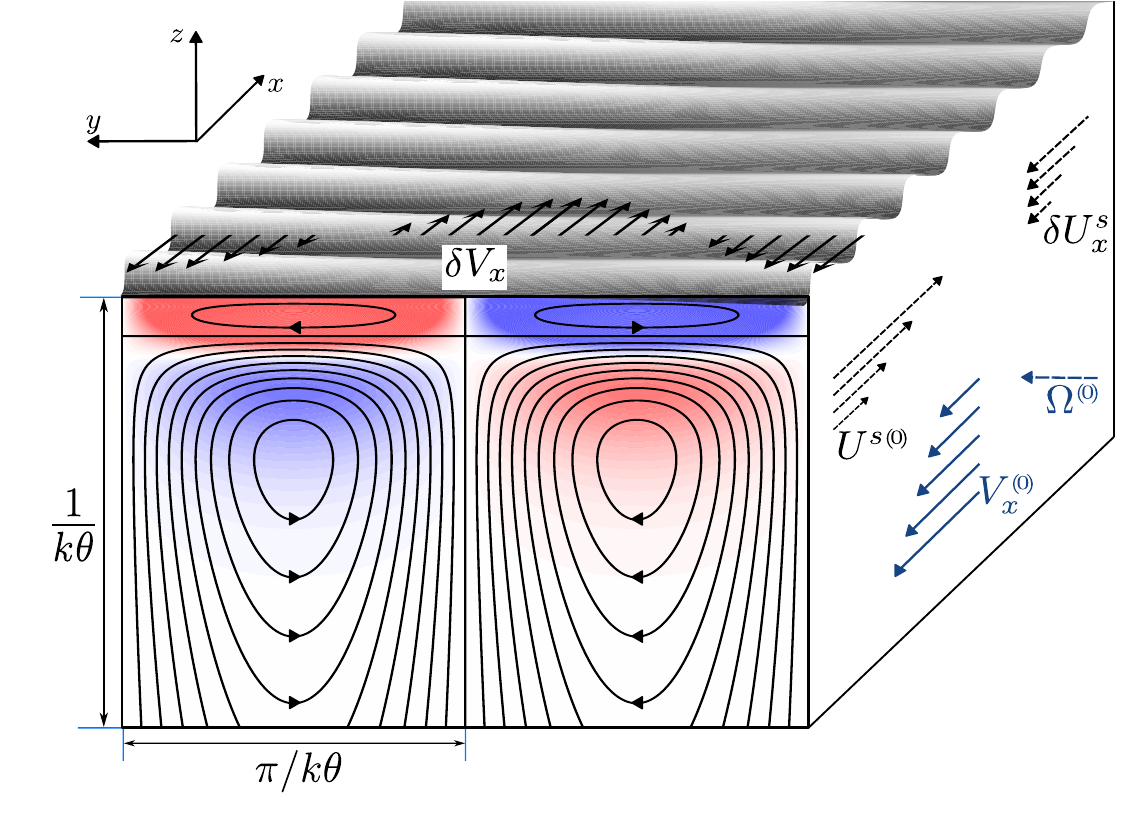}
    \caption{Sketch for the Langmuir instability development in the case when the wave scattering it taken into account. The chosen values of parameters are $\theta = 0.2\,\text{rad}$ and $\delta h / h^{\scriptscriptstyle (0)} = 0.3$. In the $Oyz$-slice, the red and blue areas correspond to the positive and negative~$\delta \Omega_x$ respectively, the areas are separated by the plane $y=0$. The solid lines are the streamlines of the circulation.}
    \label{fig:scheme}
\end{figure}

Now we impose the boundary conditions. The linearized kinematic boundary condition~\eqref{kinematic} becomes
\begin{equation}\label{kinematic_new}
    \left(\partial_t +
    V_{\alpha}\big|_{z = 0}\partial_{\alpha}\right)\zeta =
    \partial_z\phi + \zeta\partial_zV_z\big|_{z = 0},
\end{equation}
where we used the condition (\ref{upsi-cindition}) for $u^{\varpi}_z$. Because the viscosity was neglected in (\ref{u-fast-NS}), we only need one dynamical boundary condition for the pressure (\ref{dynamic_pressure}). To obtain this condition, one should express the fast oscillating pressure in terms of the potential $\phi$ and the surface elevation $\zeta$. In the case of a pure wave flow, this is realized through the Bernoulli equation, which should be linearized with respect to the wave amplitude in our approximations, so we have $p^u = - gz - \partial_t\phi$. We have to generalize the equation in the presence of the slow current, which includes the replacement of the time derivative $\partial_t$ by the substantial derivative $(\partial_t+V_i\partial_i)$. We denote by $p^{\varpi}$ the residual part of the pressure, so by definition
\begin{equation}\label{p_vortical}
    p^{u} \equiv  - gz - \left(\partial_t + V_i\partial_i\right)\phi + p^{\varpi}.
\end{equation}
Thus, the dynamic boundary condition is
\begin{equation}\label{dynamic_new}
    \left(\partial_t+V_{\alpha}\big|_{z=0}\partial_{\alpha}\right)\phi\big|_{z = 0} =
    -g\zeta+\left.p^{\varpi}\right|_{z=0},
\end{equation}
where the pressure part $p^{\varpi}$ satisfies
\begin{align}
    \label{p_vortical_eq}
        &\Delta p^{\varpi} =
        \partial_i\phi \, \Delta V_i
        +
        2 \partial_j u^{\varpi}_i \, \partial_i V_j,
    \\[5pt]
    \label{p_vortical_bc}
        &\partial_z p^{\varpi}\big|_{z=0} = \partial_{\alpha}\phi\partial_z V_\alpha,
        \quad p^{\varpi}\big|_{z\rightarrow-\infty}\rightarrow0.
\end{align}
In (\ref{p_vortical_eq}), the vortical part of the high-frequency flow ${\bm u}^\varpi$ should be restored using the equation ${\bm \varpi}^u = \mathop{\mathrm{curl}}{\bm u}^\varpi$ from (\ref{vorticity-wave}) with the boundary condition (\ref{upsi-cindition}) at the free surface and the condition ${\bm u}^\varpi\to0$ at infinity. Note that if the velocity ${\bm V}$ is purely potential, then $\Delta {\bm V}=0$ and ${\bm u}^\varpi=0$ at the right-hand side of (\ref{p_vortical_eq}), so $p^\varpi=0$ as well. The contribution $p^\varpi$ into the pressure is related to the vortical part ${\bm u}^\varpi$ of the wave motion, but as ${\bm u}^\varpi$ is a function of ${\bm V}$, the relation is not linear in ${\bm V}$.

According to~\eqref{vorticity-wave}, the ratio of the terms in the right-hand side of~\eqref{p_vortical_eq} is estimated as
\begin{equation}
    \label{p_estimate}
    \frac{2\partial_j u^{\varpi}_i \, \partial_i V_j}{\partial_i\phi \, \Delta V_i} \sim
    \frac{V}{\omega / k}.
\end{equation}
Thus, the second term in~\eqref{p_vortical_eq} should be taken into account only if the amplitude $V$ of the slow flow change in space is comparable with the phase velocity $\omega/k$ of the initial wave. Due to the above assumed restriction $V/L\ll\omega$, the limit can be  achieved only if the slow current characteristic scale is large compared to the wavelength, $kL\gg1$, so the change is accumulated in the horizontal plane. In the limit, the problem can be analysed with the ray approximation \cite{garrett1976generation} as well.

\begin{figure}[t]
    \centering
    \includegraphics[width = 1.0\linewidth]{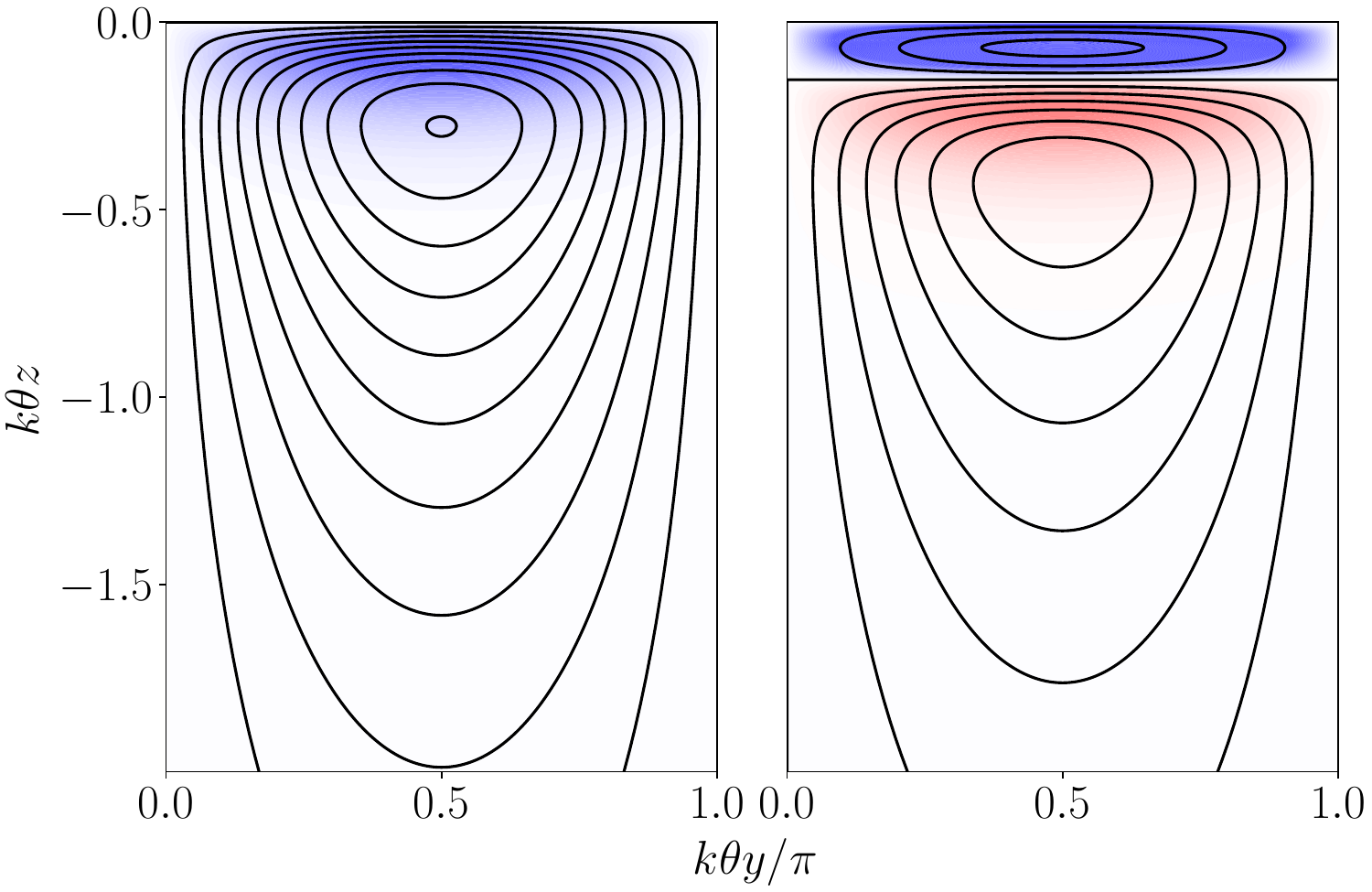}
    \caption{Sketch of the circulation in the $Oyz$-plane in the CL2-model for the first and second modes. The notations are the same as in Figure~\ref{fig:scheme}.}
    \label{fig:schemeCraik}
\end{figure}

\section{Langmuir instability mechanism}
\label{sec:Langmuir-instability}

In this Section, we consider the Langmuir circulation instability problem. We assume that the unperturbed flow is a plane monochromatic wave propagating on a coaligned vertically sheared background flow. Thus, the unperturbed flow is uniform in the spanwise direction, whereas the perturbation is modulated in this direction with a period being much larger that the wavelength. A sketch of the flow is depicted in Figure~\ref{fig:scheme}.

\subsection{The unperturbed flow}
\label{subsec:unperturbed-flow}

We assume that the wave travels in the $x$-direction in the Cartesian coordinate system. Thus, the wave potential is
\begin{equation}\label{phi_0}
    \phi^{\zero} = \Re\Big(\psi^{\zero} \exp[i\varphi^{\zero} + kz] \Big)
\end{equation}
with $\psi^{\zero}$ and $k$ being the potential amplitude and the wave number respectively and the phase $\varphi^{\zero} = kx-\omega t$. The corresponding surface elevation rate is
\begin{equation}\label{zeta_0}
    \zeta^{\zero} = \Re\Big(h^{\zero} \exp[i\varphi^{\zero}] \Big).
\end{equation}
Without loss of generality we assume $h^{\zero}>0$.According to the general scheme developed in Section \ref{subsec:wave-scattering}, the steepness of the wave is assumed to be small, $kh^{\zero}\ll1$. The slow current is shear flow, which is aligned in the same $x$-direction,
\begin{align}
    \label{V_0_general}
    \bm{V}^{\zero} =\big\{V^{\zero}_x, \, 0, \, 0\big\},
    \qquad
    \bm{\Omega}^{\zero} =\big\{0, \, \Omega^{\zero}, \, 0\big\},
\end{align}
where $\Omega^{\zero}(z) = \partial_z V^{\zero}_x(z)$. The relative slowness of the flow means that $\varepsilon \sim \Omega^{\zero}/\omega \ll 1$. One can think about the resulting motion as a generalized Gouyon wave, see e.g. Ref.~\onlinecite{abrashkin2022gerstner}. Here, we apply our general scheme to find the influence of the shear on the wave reproducing the well-known result\cite{stewart1974hf}.

We start from the Stokes drift, which $x$ component is
\begin{equation}\label{Us_0}
    U^{s\zero}_x = k\omega \big(h^{\zero}\big)^2 e^{2kz}.
\end{equation}
The vortex force~\eqref{vortex-force_new} appears to be purely potential
\begin{equation}\label{f_0}
    \bm{f}^V = k\omega \Omega^{\zero}\big(h^{\zero}\big)^2 e^{2kz} \big\{0, \, 0, \, 1 \big\}
\end{equation}
and, according to the equation~\eqref{V-slow-NS}, does not produce any contribution to the slow flow but only modifies the pressure.

\begin{figure}[t]
    \centering
    \includegraphics[width = 1.0\linewidth]{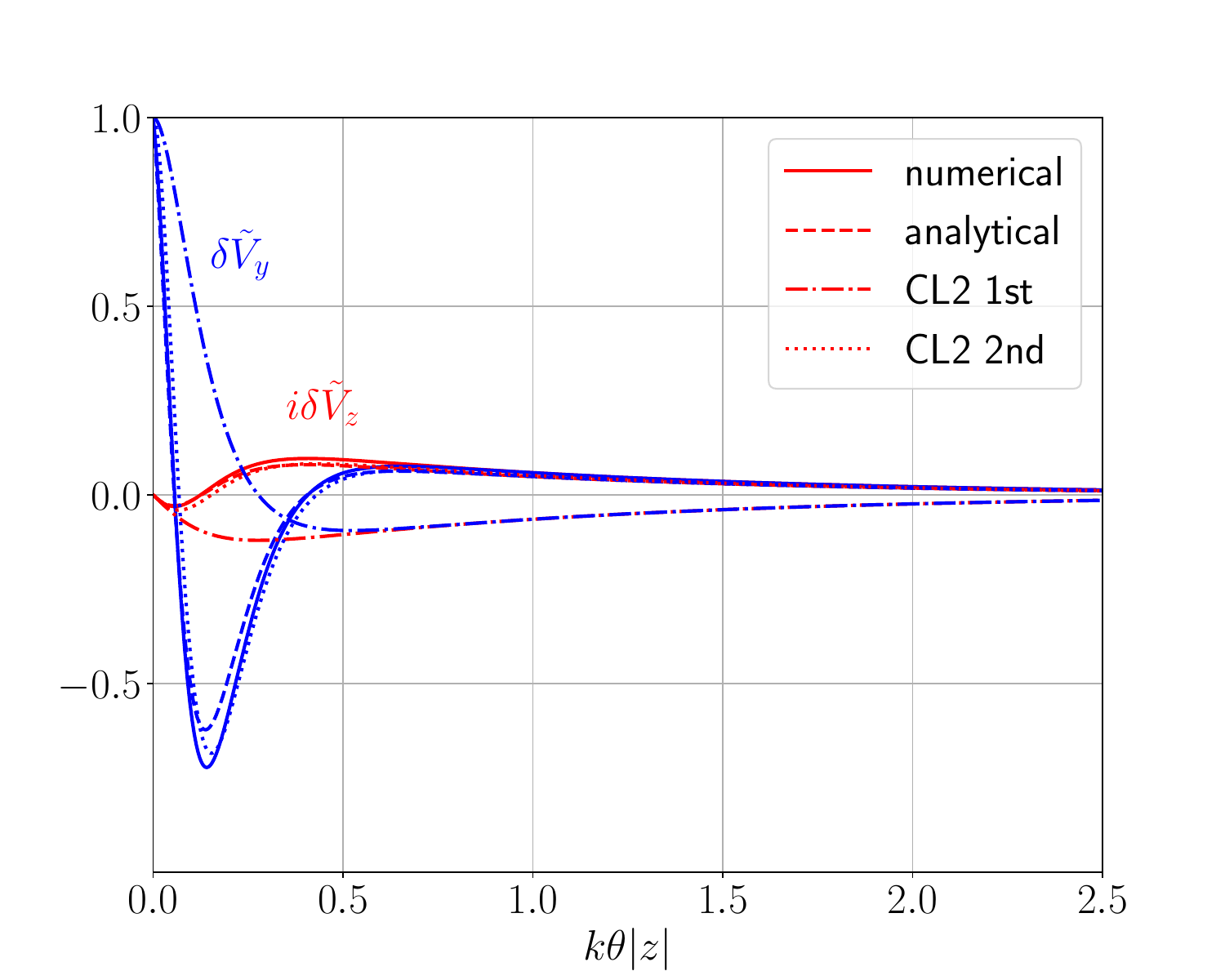}
    \caption{Vertical dependencies of $\delta\tilde V_z$ (red) and $\delta \tilde V_y$ (blue) related to $\delta \tilde \Psi$ according to (\ref{dPsi}). It was taken $\theta=0.2$. The solid lines are for numerical solution of the system of equations(\ref{dPsi_eq}) and $\delta \tilde p^\varpi=0$ (\ref{pvarpi-zero}) with $\delta\tilde p^\varpi$ calculated according to (\ref{dpressure}). The corrections of order $\lambda/\Omega^{\zero}$ were neglected. The dashed lines (almost coincide with the solid lines) correspond to the limit $\theta=0$, see (\ref{dPsi_small_theta}). The dash-dotted and dotted lines correspond to the first and second CL2 modes respectively. All the curves are normalized according to the condition $\delta \tilde V_y\vert_{z=0}=1$. The streamwise component $\delta\tilde V_x$ is proportional to $\delta\tilde V_z$, see (\ref{dV_Psi}).}
    \label{fig:dV}
\end{figure}

The dispersion relation and, thus, the phase velocity are obtained from the boundary conditions~\eqref{kinematic_new} and~\eqref{dynamic_new} on the free surface:
\begin{align}
    \label{kinematic_0_general}
    &\big(-i\omega + ikV^{\zero}_x(0)\big)h^{\zero} = k\psi^{\zero}, \\[5pt]
    \label{dynamic_0_general}
    &\big(-i\omega + ikV^{\zero}_x(0)\big)\psi^{\zero} = - gh^{\zero} + p^{\varpi\zero},
\end{align}
with the pressure
\begin{equation}
    p^{\varpi \zero} = 2ik\psi^{\zero}\Big(V^{\zero}_x(0) -
    2k\int\limits_{-\infty}^{0} dz \, e^{2kz} V^{\zero}_x(z)\Big).
\end{equation}
Here, the vortical part of the high-frequency flow ${\bm u}^\varpi$ was neglected in (\ref{p_vortical_eq}) due to the shear flow is uniform in the horizontal plane. The solution of~\eqref{kinematic_0_general}-\eqref{dynamic_0_general} linearized with respect to $\varepsilon$ is
\begin{equation}\label{dispersion_0_general}
    \omega = \sqrt{gk} + \omega^\prime,
    \qquad
    \omega^\prime = 2k^2\int\limits_{-\infty}^{0} dz \, e^{2kz} V^{\zero}_x(z),
\end{equation}
and the phase velocity $\omega/k$ appears the same as in Ref.~\onlinecite{stewart1974hf}. One can choose the reference system where $V_x^{\scriptscriptstyle(0)}\vert_{z=0}=0$. We assume that the shear flow is created by the surface stress coaligned with the $Ox$-axis, as the Langmuir instability only exists in this case \cite{leibovich1983form}. This means that $\Omega^{\scriptscriptstyle(0)}>0$ if one considers the simplest case of constant-sign shear. Under the assumptions, the correction to the wave frequency $\omega^\prime<0$. If in addition, we suppose a linear shear, ${V}^{\zero}_x = \Omega^{\zero}z$, then the correction to the wave frequency is simply $\omega^\prime = - \Omega^{\zero}/2$.

\begin{figure}[t]
    \centering
    \begin{picture}(200,180)
    \put(-30,0){
    \includegraphics[width = 1.0\linewidth]{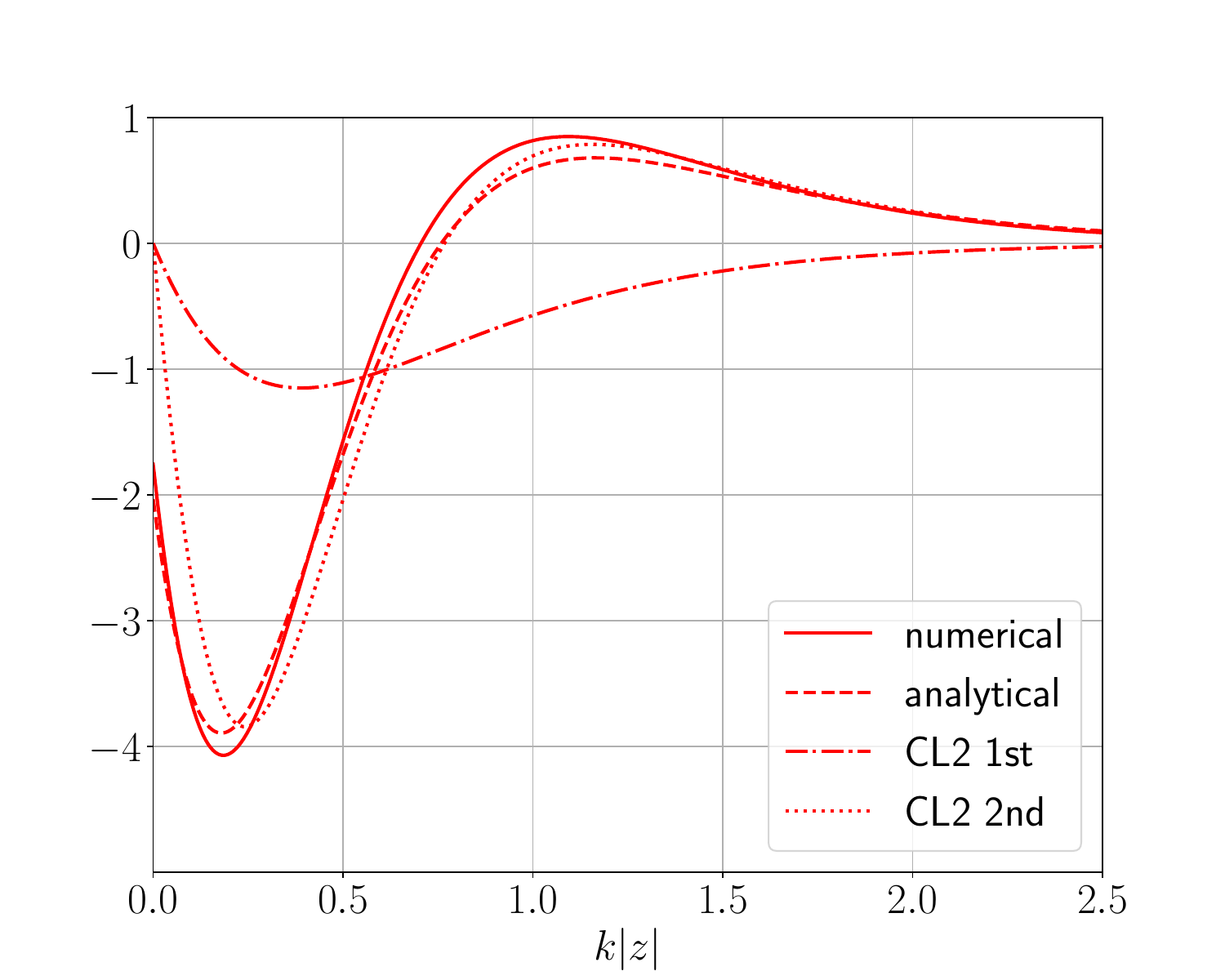}
    }
    \end{picture}
    \caption{Vertical dependencies of $\delta\tilde{\Omega}_x$ computed from $\delta\tilde{\Psi}$ according to (\ref{dPsi}). The notations and the value of $\theta$ are the same as in Figure~(\ref{fig:dV}).}
    \label{fig:dOx}
\end{figure}

\subsection{The perturbed flow}
\label{subsec:perturbation}

It is assumed in Langmuir instability CL2 mechanism~\cite{craik1977generation} that an unstable mode includes a streamwise shear flow and a circulation in vertical-spanwise plane both modulated in the spanwise direction. Here, we add into the mode structure the wave modulation, which can be thought as the result of the initial wave scattering on the modulated slow flow. The modulated wave is oblique \cite{craik1970wave} as its wave vector is tilted with regard to the $x$-direction. Hereinafter, we denote the contributions of the mode to the physical quantities with the symbol~\textquoteleft$\delta$\textquoteright, so the full fields are now
\begin{equation}
\begin{aligned}
    &\bm{V} = \bm{V}^{\zero} + \dV, \quad
    &\phi = \phi^{\zero} + \delta\phi, \\
    &\bm{\Omega} = \bm{\Omega}^{\zero} + \dO, \quad
    &\zeta = \zeta^{\zero} + \delta \zeta.
\end{aligned}
\end{equation}
The wave potential $\delta \phi$ and the surface elevation $\delta \zeta$ are:
\begin{eqnarray} \label{dphi}
    \delta\phi & = &\Re\Big(\delta\psi \, e^{\lambda t}
    \exp\big[i\varphi^{\scriptscriptstyle(1)} + kz\big]\Big),
    \\ \label{dzeta}
    \delta\zeta & = &\Re\Big(\delta h \, e^{\lambda t} \exp\big[i\varphi^{\scriptscriptstyle(1)}\big]\Big),
\end{eqnarray}
where the phase $\varphi^{\scriptscriptstyle(1)} = (k-\delta k)x + k\theta y - \omega t$ contains the spanwise modulation with wave number $k\theta$, and $\theta\ll1$ can be thought as the angle between propagation directions of the oblique and initial waves. It is assumed also that $\delta k\ll k$, so the wavenumbers of the waves are close to each other. The instability growth rate $\lambda$ is assumed to be much smaller compared to the wave frequency, $\lambda\ll\omega$, so the wave is assumed to be quasi-monochromatic. The correction to the oscillating part of the pressure can be represented in the same form (see Appendix~\ref{appendix:vortical_part_of_pressure})
\begin{equation}\label{dp_vortical}
    \delta{p}^{\varpi} = \Re\Big(\delta\tilde{p}^{\varpi}(z) \  e^{\lambda t} \exp\big[i\varphi^{\scriptscriptstyle(1)}\big]\Big).
\end{equation}
The dynamics and the spatial structure of the slow flow match with the fast-oscillation averaged dynamics of the interference between the initial and scattered waves. Thus, the time dependence is exponential, $\exp(\lambda t)$, and the spatial structure of the slow flow is determined by the phase difference $\varphi^{\scriptscriptstyle(1)} -  \varphi^{\scriptscriptstyle(0)} = k\theta y -\delta k \cdot\! x$:
\begin{align} \label{dV}
    \delta\bm{V} = \Re\Big(\delta\tilde{\bm{V}}(z) \, e^{\lambda t} \exp\big[i\big(\varphi^{\scriptscriptstyle(1)} - \varphi^{\scriptscriptstyle(0)}\big)\big]\Big).
\end{align}
The same notations are considered for $\delta{\bm \Omega}$.

\subsection{The perturbation equations}
\label{subsec:linearized-equations}

The next step is to derive the linearized equations that define the dynamics of the perturbation. The full system of equations to be linearized was described in Section~\ref{subsec:unperturbed-flow} and consists of equations (\ref{V-equation},\ref{slow-flow-BC}) for the slow flow and equations (\ref{Laplace},\ref{kinematic_new},\ref{dynamic_new},\ref{p_vortical_bc}) for the fast oscillating motion. We note that one should keep only the first summand in the right-hand side of Eq.~(\ref{p_vortical_eq}) since the perturbation amplitude is small. The slow flow part of the perturbation is completely defined by $x$-components of the velocity and vorticity $\delta \tilde V_x(z)$, $\delta\tilde\Omega_x(z)$ and its fast oscillating part is described by the oblique-wave parameters $\delta \psi$, $\delta h$, so we have to derive equations for these quantities. Following Ref.~\onlinecite{craik1977generation}, we restrict ourselves to considering the case of $x$-independent slow flow. This means that $\delta k$ in (\ref{dphi}-\ref{dV}) is negligibly small. This is a restriction which selects a definite type of unstable modes. For more broad class of modes with non-trivial $x$-dependence, a proper theoretical approach would become significantly more complicated and is beyond the scope of this consideration. Within the assumption about the homogeneity along $x$-axis, one can only state that the wavenumber $k_{\mathrm{o}}$ of the oblique wave is close to the wavenumber $k$ of the initial wave, $k_{\mathrm{o}}^2-k^2\approx k^2\theta^2 - 2k\cdot \delta k$. Here $\delta k$ should be assumed so small that it can be neglected in all the relevant equations, therefore $\delta k\lesssim \theta^2 k$. That the assumption is self-consistent is proved by the fact that the final solution for the unstable mode is nontrivial.

The approximation allows one to introduce the stream function $\dPsi$ for the circulation in the $Oyz$-plane:
\begin{align} \label{dPsi}
    &\dVy = \partial_z\dPsi,
    \quad \dVz = -\partial_y\dPsi = -ik\theta \dPsi,
    \notag\\
    &\dOx = -(\partial_z^2 - k^2\theta^2)\dPsi.
\end{align}
In what follows, we neglect all the corrections which are relatively small as $\theta^2$. However, the $\theta^2$-term should be kept in the expression for $\dOx$ (\ref{dPsi}) since $\dPsi$ below will be shown to exponentially drop when $z\to-\infty$, $\dPsi\propto \exp(k|\theta|z)$. We admit below that all the quantities describing the unstable mode are complex, that is in particular, one should drop $\mathrm{Re}$ in the definitions (\ref{phi_0},\ref{zeta_0}) for the initial wave (\ref{dphi},\ref{dzeta},\ref{dp_vortical},\ref{dV}) for the perturbation.

Let us first linearize the vorticity equation (\ref{vorticity-equation}) averaged over the fast oscillations and the slow-flow equation (\ref{V-equation}):
\begin{align}
    \label{dO_eq_init}
    &\begin{aligned}
    \big(\lambda- \nu\Delta +  V^{\zero}_x \partial_x\big)\dO
    -
    \Omega^{\zero}\big[\delta\Omega_z \bm{e}^{x}
    + \partial_y\dV\big]=\\
    =\Curl\delta\bm{f}^{\scriptscriptstyle V},
    \end{aligned}
    \\[5pt]
    \label{dV_eq_init}
    &\begin{aligned}
    \big(\lambda- \nu\Delta +  V^{\zero}_x \partial_x\big) \dV
    +
    \dVz\Omega^{\zero}\bm{e}^x
    +
    \Grad\delta\overline{P}
    =\\
    =\delta\bm{f}^{\scriptscriptstyle V},
    \end{aligned}
\end{align}
where we took into consideration the variation of the vortex force
\begin{equation} \label{df_init}
    \delta\bm{f}^{\scriptscriptstyle V} = \big[\bm{U}^{s\zero}, \, \delta\bm{\Omega}\big] +
    \big[\delta\bm{U}^{s}, \, \bm{\Omega}^{\zero}\big]
\end{equation}
as well. The variation of the Stokes drift velocity defined in~\eqref{Stokes} is
\begin{equation}
    \label{dUs_init}
    \delta\bm{U}^{s} =
    \frac{1}{2} \, \partial_l \big( \delta s_l^* \bm{u}^{\phi\zero} \big)
    + \frac{1}{2} \, \partial_l \big( s^{\zero*}_l \delta\bm{u}^{\phi}\big),
\end{equation}
where symbol ${}^\ast$ denotes the complex conjugation. In the small angle $\theta$ limit, the Stokes drift perturbation is
\begin{align}
    \label{dUs_x}
    \delta U^{s}_x ={}& 2ik^2h^{\zero}\delta\psi \, e^{\lambda t+ 2kz}
    \exp\big[i(\varphi^{\scriptscriptstyle(1)}-\varphi^{\zero}) \big],
\end{align}
and $\delta U^{s}_y = \theta\,\delta U^{s}_x$, but the $y$-component does not produce any contribution into the vortex force (\ref{df_init}). The perturbation (\ref{df_init}) of the vortex force appears to be embedded in $yz$-plane. Now, we take the $x$-components of equations (\ref{dO_eq_init},\ref{dV_eq_init}) and obtain in our approximations
\begin{align}
    \label{dO_eq_x}
    &\begin{aligned}
    \big[\lambda - \nutu \big(\partial_z^2 - k^2\theta^2\big)\big]\dtOx + 2ik^2\theta U^{s\zero}\dtVx =\\
    =ik\theta\Omega^{\scriptscriptstyle (0)}\delta\tilde{U}^{s}_x,
    \end{aligned}
    \\[5pt]
    \label{dV_eq_x}
    &\big[\lambda - \nutu \big(\partial_z^2 - k^2\theta^2\big)\big]\dtVx + \Omega^{\zero}\dtVz = 0.
\end{align}
Note that the $x$-component of the square bracket in (\ref{dO_eq_init}) was neglected as it is equal to $\partial_x \delta V_y$. Equations should be supplemented with the boundary conditions on the free surface~\eqref{slow-flow-BC}
\begin{equation}
    \label{dV_bc}
    \partial_{z}\delta V_{x}\big|_{z = 0} = \partial_{z}\delta V_{y}\big|_{z = 0} = 0,
    \quad
    \dVz\big|_{z = 0} = 0.
\end{equation}

Next, we rewrite equations (\ref{kinematic_new},\ref{dynamic_new}) substituting the expressions (\ref{phi_0},\ref{zeta_0},\ref{dphi},\ref{dzeta})
\begin{align}
    \label{kbc_delta}
    (\lambda - i\omega)\delta{h} ={}& k\delta\psi
    +
    h^{\scriptscriptstyle (0)}\partial_z\dtVz\big|_{z = 0},
    \\[5pt]
    \label{dbc_delta}
    (\lambda - i\omega)\delta\psi ={}& -g\delta{h} + \delta\tilde{p}^{\varpi}\big|_{z = 0}.
\end{align}
The pressure $\delta\tilde{p}^{\varpi}$ (\ref{dp_vortical}) evaluated on the surface can be presented in the form (see the derivation in Appendix~\ref{appendix:vortical_part_of_pressure} and the general expression for  $\delta p^\varpi$ \eqref{appendix:p_vortical_int})
\begin{align}\label{dpressure}
    \delta\tilde{p}^{\varpi}\big\vert_{z=0}
    =
    i\Omega^{\scriptscriptstyle (0)}\big\vert_{z=0}\delta{\psi}
    +
    i\psi^{\zero}\partial_z\dtVx\big|_{z = 0}
    -
    \notag \\
    -
    \psi^{\zero}\int\limits_{-\infty}^{0} dz \, e^{2kz}
    (\partial_z^2 - k^2\theta^2)\big(i\dtVx + \dtVz\big)
    .
\end{align}
For the viscid problem, the second term in (\ref{dpressure}) is nil according to the boundary conditions for the slow current~\eqref{dV_bc}. As we below consider the limit of inviscid fluid, here we keep the term.

\subsection{Solution}
\label{subsec:LC-solution}

Now we assume, that the perturbations growth rate $\lambda$ is large compared to the viscous decay rate of the perturbation slow-flow contribution. Since the Stokes drift in (\ref{dV_eq_x}) has a penetration depth $\sim 1/k$, the vertical scale of $\delta \Omega_x$ is the same depth. This means that the viscous damping rate satisfies
\begin{equation} \label{scales}
    \nutu  k^2 \ll \lambda.
\end{equation}
Below, we show that this inequality is equivalent to the limit of the small Langmuir number, $\mathrm{La}\ll\theta^2$. In this limit, the viscous scale $\delta_{\nu} \sim \sqrt{\nutu  / \lambda}$ for the slow flow is less than the wave penetration depth $1/k$, $k\delta_\nu\ll1$. As a result, $\lambda$ prevails over the viscous terms in the square brackets in (\ref{dO_eq_x},\ref{dV_eq_x}), we can then neglect the viscous term in equation~\eqref{dV_eq_x} and obtain
\begin{equation} \label{dV_Psi}
    \dtVx = -\frac{\Omega^{\zero}}{\lambda}\dtVz = \frac{ik\theta\Omega^{\zero}}{\lambda}\dtPsi
\end{equation}
due to the definition (\ref{dPsi}). According to the equalities (\ref{dV_Psi}), the mutual signs of $\dtVx$ and $\dtVz$ are fixed due to the assumption $\Omega^{\zero}>0$ and, as it will be shown below, the growth rate is real, $\lambda>0$. The downward circulation flow corresponds to the maximum amplitude of the shear flow. On the other hand, the downward flow corresponds to convergent stream lines on the surface. In nature, the center lines (which are $x$-oriented and often called `streaks') of the areas are visible to an eye due to the fact that a surface contamination and a foam are collected by the flow along the lines. The location of the downwelling zones under the maxima of the shear flow is one of the characteristic features of the Langmuir circulation, see e.g. Ref.~\onlinecite{craik_leibovich_1976}. Under the same approximation, the equation~\eqref{dO_eq_x} gives
\begin{align}
    \label{dPsi_eq}
    (k^{-2}\partial_z^2 - \theta^2)\dtPsi +
    \mu^2 \,e^{2kz}\dtPsi
    =
    \sqrt{\varepsilon}
    \, \mu \,e^{2kz}\delta{\psi},
\end{align}
where the dimensionless quantities $\mu$ and $\varepsilon$ are
\begin{equation}
    \label{mu_def}
    \mu = \dfrac{\sqrt{2\Ozero\omega}}{\lambda} \,
    k \hzero \theta,
    \quad
    \varepsilon = \dfrac{2\Ozero}{\omega^{\,}} \ll 1.
\end{equation}
As we are considering the inviscid problem, the boundary conditions for the stream function are
\begin{equation}
    \label{dPsi_bc}
    \dtPsi(0) = \dtPsi(-\infty) = 0
\end{equation}
instead of (\ref{dV_bc}). In order the wind stress be physically noticeable, the shear rate $\Omega^{\zero}$ should much exceed its value in the absence of the applied wind stress, when only the virtual wave stress produces the shear \cite{longuet1953mass}, $\Omega^{\zero} \gg (k h^{\zero})^2 \omega$. Hence, the dimensionless parameter $\varepsilon \gg (k h^{\zero})^2$ at a nonzero wind stress, that is equivalent to
\begin{equation}\label{lamOmega}
    \lambda/\Omega^{\zero} \ll \theta
\end{equation}
since $\mu$ is of order one, see below. The inequality (\ref{lamOmega}) together with the equation (\ref{dV_Psi}) means that the shear modulation amplitude is much greater than the circulation amplitude, $ \delta V_{z}/\delta V_x \ll \theta$. Below, we keep only the linear in $\theta$ corrections and drop the corrections of order of $\lambda/\Omega^{\zero}$.

It is convenient to divide the further solution process into two steps. The first step is determining the slow current part via the oblique wave parameters. On the step, we assume that the unperturbed shear flow is linear, ${V}^{\zero}_x = \Omega^{\zero}z$, that makes it possible to solve analytically (\ref{dPsi_eq}). Note that the parameters $\mu$ and $\varepsilon$ are $z$-independent after $\Ozero$. On the second step, we define the instability growth rate $\lambda$ from the requirement that the free surface equations~\eqref{kinematic_new} and~\eqref{dynamic_new} have nontrivial solution.

The analytical solution of the equation~\eqref{dPsi_eq} in the case of the linear initial shear flow is:
\begin{align} \label{dPsi_small_theta}
    \dtPsi = \dfrac{\sqrt{\varepsilon}}{\mu}
    \Bigg[1 -
    \left(1 + \dfrac{1 - {\mathrm J}_0(\mu)}{{\mathrm J}_{0}(\mu)}\,e^{|\theta|k z}\right)
    {\mathrm J}_{0}\big(\mu \, e^{kz}\big)\Bigg]\delta{\psi}.
\end{align}
Expression (\ref{dPsi_small_theta}) is not exact, the exact solution owns small corrections of order $\theta$. With such an accuracy, one can approximate the exponent $e^{|\theta|k z}\to1$ when $kz\sim 1$ so the square bracket in (\ref{dPsi_small_theta}) is equal to $1-{\mathrm J}_{0}\big(\mu \, e^{kz}\big)/{\mathrm J}_{0}(\mu)$. In particular, $\partial_z\delta\tilde\Psi\vert_{z=0} = k\sqrt{\varepsilon}\,{\mathrm J}_1(\mu)/{\mathrm J}_0(\mu)$, which is present in the last term in (\ref{kbc_delta}) and in the second term in (\ref{dpressure}). The horizontal component of circulation velocity is much greater than the vertical one at $kz\sim1$, $\delta V_z/\delta V_y\sim\theta$. At large depths $|z|\gtrsim1/k|\theta|$, the stream function decays exponentially as the square bracket in (\ref{dPsi_small_theta}) is $(1-1/{\mathrm J}_0(\mu))  \, e^{|\theta|k z}$, so the circulation velocity components have equal amplitudes, $\tilde V_z = i\mathop{\mathrm{sign}}\theta\cdot \delta \tilde V_y$. Thus, the circulation penetration depth is $O(1/k|\theta|)$. However, the vorticity related to the stream function has a penetration depth only $O(1/k)$ according to (\ref{dPsi}),
\begin{equation} \label{dOx_solution}
    \delta\tilde \Omega_x = i\sqrt{\varepsilon} \,
    \dfrac{\mu{\mathrm J}_{0}(\mu \, e^{kz})}{{\mathrm J}_0(\mu)}
    \, e^{2kz} \omega k\, \delta{h},
\end{equation}
where we used the approximate equality $\delta \psi = -i (\omega/k)\delta h$. Note that in the CL2 model, the vertical dependence of the vorticity is described by formally the same function, $\delta\tilde\Omega_x = (\mu k)^2 e^{2kz}{\mathrm J}_0(\mu e^{kz})$. However, one should keep in mind that $\mu$ has different values in the CL2 and in our model, see below.

Equations (\ref{dV_Psi},\ref{dPsi_small_theta},\ref{dOx_solution}) express the slow flow via the oblique wave parameters $\delta h$, $\delta\psi$. To be able to write explicitly equations (\ref{kbc_delta},\ref{dbc_delta}) in terms of these parameters, we need to do this, first for the residual part of the pressure $\delta{p}^{\varpi}$ using~\eqref{dpressure} in the inviscid limit:
\begin{eqnarray}
    \label{dpressure_small_theta}
    \delta\tilde{p}^{\varpi}\big|_{z = 0}
    & = &
    iF(\mu)\,\Omega^{\zero}\delta\psi,
\end{eqnarray}
where the function
\begin{equation}
    F(\mu) = \dfrac{4}{\mu}\dfrac{{\mathrm J}_{1}(\mu)}{{\mathrm J}_{0}(\mu)} - 1
\end{equation}
is even in $\mu$. In (\ref{dpressure_small_theta}), we neglected corrections of order of $\lambda/\Omega^{\zero}$ in accordance with (\ref{lamOmega}), the corrections stem from the term having $\delta \tilde V_z$ in (\ref{dpressure}). Now, the linear system of equations (\ref{kbc_delta},\ref{dbc_delta}), solving for $\delta h$, $\delta\psi$, has a nontrivial solution if 
\begin{equation}\label{pvarpi-zero}
    \delta \tilde p^\varpi \vert_{z=0} = i \Omega^{\zero}\,\delta\psi,
    \qquad \text{or} \qquad
    F(\mu) = 1,
\end{equation}
where we have accounted for the frequency shift (\ref{dispersion_0_general}) for the fundamental wave and the corrections of order of $\lambda/\Omega^{\zero}$ stemming again from $\delta \tilde V_z$-term in (\ref{kbc_delta}) were neglected.  Among all the existing solutions for (\ref{pvarpi-zero}), we should choose that corresponding to the maximum positive real part of $\lambda$. The numerical solution is then
\begin{equation}\label{the_mu}
    \mu_\star \simeq 5.14,
    \quad
    \lambda =
    \dfrac{\sqrt{2\Omega^{\scriptscriptstyle{(0)}}\omega}}
    {\mu_\star}
    \, k h^{\scriptscriptstyle{(0)}} |\theta|.
\end{equation}
For finite $\theta$, the numerical solution of $\delta \tilde p^\varpi/\delta\psi=i \Omega^{\zero}$ (\ref{pvarpi-zero}) with $\delta\tilde p^\varpi$ and $\delta \tilde\Psi$ calculated according to (\ref{dpressure},\ref{dPsi_eq}) respectively gives $\mu(\theta) \simeq \mu_\star + 2.15 |\theta|$.

Let us test the properties of the solution with respect to the transformation $\theta\to-\theta$, that leads to $\mu\to-\mu$ as well. Without loss of generality, we can admit that $\delta h $ is pure real and negative $\delta h < 0$. The phase difference between the initial and oblique waves along the streamwise direction is small since $\delta k\,x\ll1$ for $kx=O(1)$, so the phase difference $\varphi^{\scriptscriptstyle(1)} - \varphi^{\scriptscriptstyle(0)} = k\theta y$ in (\ref{dV}). Then both $\delta \tilde V_z$ (\ref{dPsi}) and $\delta \tilde V_x$ (\ref{dV_Psi}) are purely real and even in $\theta$, so $\delta V_z \sim \cos(k\theta y)$ (the signs of the left and right hand sides of the similarity coincide) is an even function of $\theta$ after $\delta \tilde V_x$, whereas $\delta \tilde V_y$ is purely imaginary and odd in $\theta$, so $\delta V_y \sim -\theta \sin(k\theta y)$ (the sign is taken at the surface) is an even function of $\theta$ as well. This means that the unstable mode is twice degenerate; two possible configurations have an identical slow flow spatial structure and oblique waves with opposite spanwise wave numbers.

\section{Discussion}
\label{sec:discussion}

Knowing that $\mu\sim1$, let us rewrite the inequality (\ref{scales}) which ensures the inviscid limit in a more known form. We introduce the friction velocity $u_\ast$ according to the definition $\Omega^{\scriptscriptstyle{(0)}} = u_\ast^2/\nu_{\scriptscriptstyle T}$. Here we have assumed that there is some small-scale and relatively fast turbulent flow on the background of the slow flow ${\bm V} = {\bm V}^{\zero} + \delta {\bm V}$, which affects the slow flow via turbulent viscosity $\nu_{\scriptscriptstyle T}$. In particular, the molecular diffusion in the equations (\ref{V-slow-NS},\ref{V-equation},\ref{dO_eq_init},\ref{dV_eq_init},\ref{dO_eq_x},\ref{dV_eq_x}) and the inequality (\ref{scales}) should be substituted by the turbulent one. Note that at the same time, the fast dynamics of the small scales may be relatively slow compared to the dynamics of the surface wave flow ${\bm u}$. In the case, one should use molecular diffusion in equations (\ref{vorticity-equation},\ref{vorticity-equation-lin}) for the high-frequency component assuming that ${\bm V}$ is the full slow vortical flow, which includes small scales as well. However, the interaction between the wave motion and the small-scale turbulence is anticipated to be weak. Then, the inequality (\ref{scales}) means that the Langmuir number
\begin{equation}\label{La}
    \mathrm{La}
    \equiv
    \frac{\nu_{\scriptscriptstyle T}k^2}{k h^{\zero}\sqrt{\omega \Omega^{\zero}}}
    =
    \frac{\nu_{\scriptscriptstyle T}^{3/2} k}{ h^{\scriptscriptstyle(0)} u_\ast\sqrt{\omega}}
\end{equation}
is far smaller than $\theta$, $\mathrm{La}\ll|\theta|$.

Now we analyze the obtained results and compare them with the results of the CL2-model\cite{craik1977generation}. Note that we come down to the CL2-model equation~\cite{craik1977generation} if we neglect the variation of Stokes drift, i.e. if we let the right-hand side of~\eqref{dPsi_eq} be equal to zero. In this case, the eigenvalue problem  (\ref{dPsi_eq},\ref{dPsi_bc}) leads to the circulation mode structure $\delta\tilde\Psi = {\mathrm J}_{\theta}(\mu\,e^{kz})$ and the first boundary condition in (\ref{dPsi_bc}) gives ${\mathrm J}_{\theta}(\mu)=0$, where $\mathrm{J}_n(\mu)$ is the $n$-th order Bessel function. Thus, the perturbation grows rate in the CL2-model corresponds to the lowest root $\mu_{\scriptscriptstyle\mathrm{CL}2,1}\simeq 2.4$ in the limit $\theta\ll1$. The next root is $\mu_{{\scriptscriptstyle\mathrm{CL}2,2}}\simeq 5.5$, which correspond to the second mode having the greatest growth rate after the first mode.

The derived spatial modal structure is schematically plotted in Figure~\ref{fig:scheme}. One can see that the circulation consists of two vortices of opposite signs one above the other. The vertical dimension of the upper vortex is of order of $1/k$ and the bottom vortex penetration depth is $\sim 1/k\theta$. For comparison, the structures of the first and second unstable modes with the greatest growth rates in the CL2-model are depicted in Figure~\ref{fig:schemeCraik}. In a qualitative sense, the general vertical structure of the circulation flow in Figure~\ref{fig:scheme} corresponds to the second mode in CL2-model. In particular, the numerical value $\mu_{{\scriptscriptstyle\mathrm{CL}2,2}}$ is close to that found in our model $\mu_\star$ (\ref{the_mu}). The detailed structure of the unstable mode in our model is deformed by the scattered wave compared to the structure of the second CL2-mode. The vertical dependencies for all the modes are plotted in Figures~\ref{fig:dV} and \ref{fig:dOx}. Qualitatively, one can say that the wave scattering process stops the growth of the first mode in CL2-model and accelerates the growth of the second mode since $\mu_\star < \mu_{{\scriptscriptstyle\mathrm{CL}2,2}}$ and the growth rate $\lambda$ is inversely proportional to $\mu$, see (\ref{the_mu}).

The presented analysis of the Langmuir instability assumes that the flow has an exactly space periodic structure. In reality, the periodicity should be lost at some length $L_c$. For the analysis to be valid, the length should be greater than the distance over which the wave propagates during the characteristic time $1/\lambda$ along the streamwise direction, which is $L_{c,x} > \omega/(k\lambda)$. Taking into account (\ref{the_mu},\ref{mu_def}), one can rewrite the inequality in the form $L_{c,x} > k^{-1}/(\sqrt{\varepsilon}\,\theta \, kh^{\scriptscriptstyle (0)})$,  which far exceeds the spanwise wavelength. In the spanwise direction, the periodicity should be kept at smaller length $L_{c,y} \sim \mathop{\mathrm{max}}(\theta L_{c,x},\,1/k\theta)$. If the periodicity appears to be disrupted at shorter distances, then the oblique wave is not correlated with the Langmuir circulation and one should drop it out during the Langmuir instability analysis. As a result, one arrives back to the CL2-model\cite{craik1977generation}, where only $x$-independence of the flow at distances $1/\delta k$ much greater than $1/k\theta$ is assumed.

The largest growth rate $\lambda$ (\ref{the_mu}) corresponds to $\theta\sim1$. Nevertheless, an established Langmuir circulation is often characterized by a small angle $\theta\ll1$\cite{leibovich2009langmuir}. The final theta selection is produced by the hydrodynamics nonlinearity or the fact that the wind-induced shear is slightly deviated from the wave propagation direction. For example, observations in ocean\cite{smith1992observed} detected the initial wave having the wavelength of \hbox{$\sim\!25\,\text{m}$} and the wave steepness \hbox{$k h^\zero \sim0.05$}. After the air wind is established, a Langmuir circulation with spanwise period equal to $2/3$ of the wavelength developed within $13\,\text{s}$. During the next hour, the period increased up to $60\,\text{m}$, that corresponds to $\theta \sim 0.4$. When the perturbation mode achieves a finite amplitude, the oblique and initial waves form together a \textquotedblleft crossing sea state\textquotedblright. The modulational instability for this state is characterized by the growth rates which are higher than those for one-dimensional wave flow\cite{manna2021modulational}, see also a brief review in Ref.~\onlinecite{yin2023modeling}. This again shows that the three-dimensionality of wave flow is essential for a proper description of flow in  the ocean surface boundary layer. Considering an extension of the scope of our approach, we believe that the developed mathematical scheme should be appropriate for the analysis of wave-current interaction in flows confined in a basin, see e.g. Refs.~\onlinecite{filatov2018experimental,filatov2022generation}. The confinement preserves the correlation between the vortical flow and the surface waves, which are mostly standing ones in this case. At the initial stage, when only surface waves are excited, the leading mechanism of the interaction is a virtual wave stress \cite{filatov2016nonlinear,filatov2022generation}, which can be enhanced due to the presence of a surface contamination in the form of a liquid elastic film \cite{parfenyev2019formation}. However, during the development of the large-scale vortical flow, the oblique wave appearance shows evidence of the wave scattering by the vortical flow \cite{filatov2018experimental}. The presented approach, combined with a more detailed analysis of experimental data, can answer whether a wave-current interaction loop plays any role in the formation of a large-scale flow, along with other mechanisms, including the two-dimensional inverse energy cascade \cite{colombi2022coexistence}.

\section{Acknowledgments}

The work was supported by Russian Science Foundation (project no. 23-72-30006).

\section{Author declarations}

\noindent \textbf{Conflict of Interest}

The authors have no conflicts to disclose.

\section{Data availability}

The data that support the findings of this study are available from the corresponding author upon reasonable request.

\onecolumngrid

\appendix

\section{Derivation of the pressure related to the high-frequency part of the vorticity}
\label{appendix:vortical_part_of_pressure}

The equation of the oscillating part of the pressure arises from the Navier-Stokes equation for the wave motion
\begin{equation}\label{appendix_ns_phi_varpi}
    \partial_t \bm{u}^{\phi} + \left(\bm{u}^{\phi}\nabla\right)\bm{V} + \left(\bm{V}\nabla\right)\bm{u}^{\phi} +
    \partial_t \bm{u}^{\varpi} + \left(\bm{u}^{\varpi}\nabla\right)\bm{V} + \left(\bm{V}\nabla\right)\bm{u}^{\varpi} =
    -\nabla p^{u} + \bm{g}.
\end{equation}
Let us group the potential and vortical terms together
\begin{equation}\label{appendix:ns_phi_varpi_group}
    \Bigl(\partial_t \bm{u}^{\varpi} + \left(\bm{u}^{\varpi}\nabla\right)\bm{V} + \left(\bm{V}\nabla\right)\bm{u}^{\varpi}\Bigr)
    -
     \left[\bm{u}^{\phi},\bm{\Omega}\right]
     =
    -\nabla\Bigl(p^{u} + gz + \left(\partial_t+ V_i\partial_i\right)\phi\Bigr),
\end{equation}
using
$$
    \left(\bm{u}\nabla\right)\bm{V} + \left(\bm{V}\nabla\right)\bm{u} =
    \nabla\left(\bm{u}\bm{V}\right)
    -
    \left[\bm{V},\bm{\omega}\right]
    -
    \left[\bm{u},\bm{\Omega}\right],
$$
valid for the incompressible velocity field. The terms inside the gradient at the right-hand side of~\eqref{appendix:ns_phi_varpi_group} are interpreted as the vortical pressure part and denoted as ``$p^\varpi$”,
\begin{align}
    \label{appendix:p_vortical}
    &p^{\varpi}\equiv p^{u} + gz + \left(\partial_t+ V_i\partial_i\right)\phi,
    \\[5pt]
    \label{appendix:p_vortical_eq}
    &\Delta p^{\varpi}
    =
    f^{\varpi},
    \qquad
    f^{\varpi}
    =
    \partial_i\phi \, \Delta V_i + 2 \partial_j u^{\varpi}_i \, \partial_i V_j .
\end{align}
The boundary condition for $p^\varpi$ on the free surface arises from the z-component of
equation \eqref{appendix:ns_phi_varpi_group}. Taking into account the relative smallness of ${\bm u}^\varpi$ (\ref{u-division}), the boundary conditions $u^\varpi_z\vert_{z=0}=0$ (\ref{upsi-cindition}) and $V_z\vert_{z=0}=0$ (\ref{slow-flow-BC}), it follows
\begin{equation}
    \label{appendix:p_vortical_bc}
    \left.\partial_z p^{\varpi}\right|_{z=0}
    =
    \partial_\alpha \phi \, \partial_z V_\alpha\big\vert_{z=0},
    \qquad
    \left.p^{\varpi}\right|_{z\rightarrow-\infty}\rightarrow 0.
\end{equation}
The differential problem~\eqref{appendix:p_vortical_eq}, \eqref{appendix:p_vortical_bc} can be analytically solved in Fourier space described by the two-dimensional wavenumber $q$ (applying the two-dimensional Fourier transform on functions of the $(x,y$) plane)
\begin{equation}
    \left(\partial^2_z-q^2\right)\dpressure_{\bf q}=f^{\varpi}_{\bf q},
    \qquad
    \dpressure_{\bf q}
    =
   - \dfrac{e^{qz}}{2q}\int\limits_{-\infty}^{0} dz' e^{qz'}f^{\varpi}_{\bf q}(z')
    -
    \int\limits_{-\infty}^{0}dz' \dfrac{e^{-q|z-z'|}}{2q}f^{\varpi}_{\bf q}(z')
    +
    \frac{e^{qz}}{q}\big(\partial_\alpha \phi\partial_z V_{\alpha}\big\vert_{z=0}\big)_{\bf q}.
\end{equation}

Now, we turn to calculations in the sake of Langmuir instability analysis. We get $ f^{\varpi} = \partial_i\phi \, \Delta V_i$ in (\ref{appendix:p_vortical_eq}), since the amplitude of the perturbed flow is small, see~\eqref{p_estimate}. Thus, we deduce
\begin{equation}
    \label{appendix:p_vortical_int}
    \dpressure_{\bf q}\big|_{z=0} =
    - \dfrac{1}{q}\int\limits_{-\infty}^{0} dz\, e^{qz} \big(\partial_i\phi \Delta V_{i}\big)_{\bf q} +
    \dfrac{1}{q}
    \big(\partial_\alpha \phi\, \partial_z V_{\alpha}\big\vert_{z=0}\big)_{\bf q}.
\end{equation}
The equality~\eqref{dp_vortical} can be proved by the substitution of~\eqref{dphi} and~\eqref{dV} in~\eqref{appendix:p_vortical_int}, and further by direct calculations, we show that only the harmonics oscillating as $\exp[i{\bf q}\cdot\!{\bf x}] = \exp[i(k-\delta k)x + iky\theta]$ yields a nonzero contribution.

\bigskip
\twocolumngrid

\begin{thebibliography}{33}%
\makeatletter
\providecommand \@ifxundefined [1]{%
 \@ifx{#1\undefined}
}%
\providecommand \@ifnum [1]{%
 \ifnum #1\expandafter \@firstoftwo
 \else \expandafter \@secondoftwo
 \fi
}%
\providecommand \@ifx [1]{%
 \ifx #1\expandafter \@firstoftwo
 \else \expandafter \@secondoftwo
 \fi
}%
\providecommand \natexlab [1]{#1}%
\providecommand \enquote  [1]{``#1''}%
\providecommand \bibnamefont  [1]{#1}%
\providecommand \bibfnamefont [1]{#1}%
\providecommand \citenamefont [1]{#1}%
\providecommand \href@noop [0]{\@secondoftwo}%
\providecommand \href [0]{\begingroup \@sanitize@url \@href}%
\providecommand \@href[1]{\@@startlink{#1}\@@href}%
\providecommand \@@href[1]{\endgroup#1\@@endlink}%
\providecommand \@sanitize@url [0]{\catcode `\\12\catcode `\$12\catcode
  `\&12\catcode `\#12\catcode `\^12\catcode `\_12\catcode `\%12\relax}%
\providecommand \@@startlink[1]{}%
\providecommand \@@endlink[0]{}%
\providecommand \url  [0]{\begingroup\@sanitize@url \@url }%
\providecommand \@url [1]{\endgroup\@href {#1}{\urlprefix }}%
\providecommand \urlprefix  [0]{URL }%
\providecommand \Eprint [0]{\href }%
\providecommand \doibase [0]{https://doi.org/}%
\providecommand \selectlanguage [0]{\@gobble}%
\providecommand \bibinfo  [0]{\@secondoftwo}%
\providecommand \bibfield  [0]{\@secondoftwo}%
\providecommand \translation [1]{[#1]}%
\providecommand \BibitemOpen [0]{}%
\providecommand \bibitemStop [0]{}%
\providecommand \bibitemNoStop [0]{.\EOS\space}%
\providecommand \EOS [0]{\spacefactor3000\relax}%
\providecommand \BibitemShut  [1]{\csname bibitem#1\endcsname}%
\let\auto@bib@innerbib\@empty
\bibitem [{\citenamefont {Hamlington}\ \emph {et~al.}(2014)\citenamefont
  {Hamlington}, \citenamefont {Van~Roekel}, \citenamefont {Fox-Kemper},
  \citenamefont {Julien},\ and\ \citenamefont
  {Chini}}]{hamlington2014langmuir}%
  \BibitemOpen
  \bibfield  {author} {\bibinfo {author} {\bibfnamefont {P.~E.}\ \bibnamefont
  {Hamlington}}, \bibinfo {author} {\bibfnamefont {L.~P.}\ \bibnamefont
  {Van~Roekel}}, \bibinfo {author} {\bibfnamefont {B.}~\bibnamefont
  {Fox-Kemper}}, \bibinfo {author} {\bibfnamefont {K.}~\bibnamefont {Julien}},\
  and\ \bibinfo {author} {\bibfnamefont {G.~P.}\ \bibnamefont {Chini}},\
  }\bibfield  {title} {\enquote {\bibinfo {title} {{Langmuir}--submesoscale
  interactions: Descriptive analysis of multiscale frontal spindown
  simulations},}\ }\href@noop {} {\bibfield  {journal} {\bibinfo  {journal}
  {Journal of Physical Oceanography}\ }\textbf {\bibinfo {volume} {44}},\
  \bibinfo {pages} {2249--2272} (\bibinfo {year} {2014})}\BibitemShut {NoStop}%
\bibitem [{\citenamefont {Thorpe}(2004)}]{thorpe2004langmuir}%
  \BibitemOpen
  \bibfield  {author} {\bibinfo {author} {\bibfnamefont {S.~A.}\ \bibnamefont
  {Thorpe}},\ }\bibfield  {title} {\enquote {\bibinfo {title} {{Langmuir}
  circulation},}\ }\href@noop {} {\bibfield  {journal} {\bibinfo  {journal}
  {Annual Review of Fluid Mechanics}\ }\textbf {\bibinfo {volume} {36}},\
  \bibinfo {pages} {55--79} (\bibinfo {year} {2004})}\BibitemShut {NoStop}%
\bibitem [{\citenamefont {Teixeira}(2019)}]{teixeira2019}%
  \BibitemOpen
  \bibfield  {author} {\bibinfo {author} {\bibfnamefont {M.~A.~C.}\
  \bibnamefont {Teixeira}},\ }\bibfield  {title} {\enquote {\bibinfo {title}
  {{Langmuir} circulation and instability},}\ }in\ \href@noop {} {\emph
  {\bibinfo {booktitle} {Encyclopedia of Ocean Sciences (Third Edition)}}},\
  \bibinfo {editor} {edited by\ \bibinfo {editor} {\bibfnamefont {J.~K.}\
  \bibnamefont {Cochran}}, \bibinfo {editor} {\bibfnamefont {H.~J.}\
  \bibnamefont {Bokuniewicz}},\ and\ \bibinfo {editor} {\bibfnamefont {P.~L.}\
  \bibnamefont {Yager}}}\ (\bibinfo  {publisher} {Academic Press},\ \bibinfo
  {address} {Oxford},\ \bibinfo {year} {2019})\ pp.\ \bibinfo {pages}
  {92--106}\BibitemShut {NoStop}%
\bibitem [{\citenamefont {Weller}\ and\ \citenamefont
  {Price}(1988)}]{WELLER1988711}%
  \BibitemOpen
  \bibfield  {author} {\bibinfo {author} {\bibfnamefont {R.~A.}\ \bibnamefont
  {Weller}}\ and\ \bibinfo {author} {\bibfnamefont {J.~F.}\ \bibnamefont
  {Price}},\ }\bibfield  {title} {\enquote {\bibinfo {title} {{Langmuir}
  circulation within the oceanic mixed layer},}\ }\href@noop {} {\bibfield
  {journal} {\bibinfo  {journal} {Deep Sea Research Part A. Oceanographic
  Research Papers}\ }\textbf {\bibinfo {volume} {35}},\ \bibinfo {pages}
  {711--747} (\bibinfo {year} {1988})}\BibitemShut {NoStop}%
\bibitem [{\citenamefont {Chang}\ \emph {et~al.}(2019)\citenamefont {Chang},
  \citenamefont {Huntley}, \citenamefont {Kirwan}, \citenamefont {Carlson},
  \citenamefont {Mensa}, \citenamefont {M.}, \citenamefont {Novelli},
  \citenamefont {{\"O}zg{\"o}kmen}, \citenamefont {Fox-Kemper}, \citenamefont
  {Pearson}, \citenamefont {Pearson}, \citenamefont {Harcourt},\ and\
  \citenamefont {C.}}]{chang2019small}%
  \BibitemOpen
  \bibfield  {author} {\bibinfo {author} {\bibfnamefont {H.}~\bibnamefont
  {Chang}}, \bibinfo {author} {\bibfnamefont {H.~S.}\ \bibnamefont {Huntley}},
  \bibinfo {author} {\bibfnamefont {A.}~\bibnamefont {Kirwan}}, \bibinfo
  {author} {\bibfnamefont {D.~F.}\ \bibnamefont {Carlson}}, \bibinfo {author}
  {\bibfnamefont {J.~A.}\ \bibnamefont {Mensa}}, \bibinfo {author}
  {\bibfnamefont {S.}~\bibnamefont {M.}}, \bibinfo {author} {\bibfnamefont
  {G.}~\bibnamefont {Novelli}}, \bibinfo {author} {\bibfnamefont {T.~M.}\
  \bibnamefont {{\"O}zg{\"o}kmen}}, \bibinfo {author} {\bibfnamefont
  {B.}~\bibnamefont {Fox-Kemper}}, \bibinfo {author} {\bibfnamefont
  {B.}~\bibnamefont {Pearson}}, \bibinfo {author} {\bibfnamefont
  {J.}~\bibnamefont {Pearson}}, \bibinfo {author} {\bibfnamefont {R.~R.}\
  \bibnamefont {Harcourt}},\ and\ \bibinfo {author} {\bibfnamefont {P.~A.}\
  \bibnamefont {C.}},\ }\bibfield  {title} {\enquote {\bibinfo {title}
  {Small-scale dispersion in the presence of {Langmuir} circulation},}\
  }\href@noop {} {\bibfield  {journal} {\bibinfo  {journal} {Journal of
  Physical Oceanography}\ }\textbf {\bibinfo {volume} {49}},\ \bibinfo {pages}
  {3069--3085} (\bibinfo {year} {2019})}\BibitemShut {NoStop}%
\bibitem [{\citenamefont {Craik}(1977)}]{craik1977generation}%
  \BibitemOpen
  \bibfield  {author} {\bibinfo {author} {\bibfnamefont {A.~D.~D.}\
  \bibnamefont {Craik}},\ }\bibfield  {title} {\enquote {\bibinfo {title} {The
  generation of {Langmuir} circulations by an instability mechanism},}\
  }\href@noop {} {\bibfield  {journal} {\bibinfo  {journal} {Journal of Fluid
  Mechanics}\ }\textbf {\bibinfo {volume} {81}},\ \bibinfo {pages} {209--223}
  (\bibinfo {year} {1977})}\BibitemShut {NoStop}%
\bibitem [{\citenamefont {Leibovich}(2009)}]{leibovich2009langmuir}%
  \BibitemOpen
  \bibfield  {author} {\bibinfo {author} {\bibfnamefont {S.}~\bibnamefont
  {Leibovich}},\ }\bibfield  {title} {\enquote {\bibinfo {title} {{Langmuir}
  circulation and instability},}\ }\href@noop {} {\bibfield  {journal}
  {\bibinfo  {journal} {Elements of Physical Oceanography: A derivative of the
  Encyclopedia of Ocean Sciences}\ }\textbf {\bibinfo {volume} {124}},\
  \bibinfo {pages} {288} (\bibinfo {year} {2009})}\BibitemShut {NoStop}%
\bibitem [{\citenamefont {Faller}\ and\ \citenamefont
  {Caponi}(1978)}]{faller1978laboratory}%
  \BibitemOpen
  \bibfield  {author} {\bibinfo {author} {\bibfnamefont {A.~J.}\ \bibnamefont
  {Faller}}\ and\ \bibinfo {author} {\bibfnamefont {E.~A.}\ \bibnamefont
  {Caponi}},\ }\bibfield  {title} {\enquote {\bibinfo {title} {Laboratory
  studies of wind-driven {Langmuir} circulations},}\ }\href@noop {} {\bibfield
  {journal} {\bibinfo  {journal} {Journal of Geophysical Research: Oceans}\
  }\textbf {\bibinfo {volume} {83}},\ \bibinfo {pages} {3617--3633} (\bibinfo
  {year} {1978})}\BibitemShut {NoStop}%
  \bibitem{smith1992observed}
  J.~A. Smith, ``Observed growth of Langmuir circulation,'' Journal of Geophysical Research: Oceans, \textbf{97}(C4), 5651 (1992)\BibitemShut {NoStop}%
\bibitem [{\citenamefont {Sullivan}\ and\ \citenamefont
  {McWilliams}(2019)}]{sullivan_mcwilliams_2019}%
  \BibitemOpen
  \bibfield  {author} {\bibinfo {author} {\bibfnamefont {P.~P.}\ \bibnamefont
  {Sullivan}}\ and\ \bibinfo {author} {\bibfnamefont {J.~C.}\ \bibnamefont
  {McWilliams}},\ }\bibfield  {title} {\enquote {\bibinfo {title} {{Langmuir}
  turbulence and filament frontogenesis in the oceanic surface boundary
  layer},}\ }\href@noop {} {\bibfield  {journal} {\bibinfo  {journal} {Journal
  of Fluid Mechanics}\ }\textbf {\bibinfo {volume} {879}},\ \bibinfo {pages}
  {512–553} (\bibinfo {year} {2019})}\BibitemShut {NoStop}%
\bibitem [{\citenamefont {Craik}\ and\ \citenamefont
  {Leibovich}(1976)}]{craik_leibovich_1976}%
  \BibitemOpen
  \bibfield  {author} {\bibinfo {author} {\bibfnamefont {A.~D.~D.}\
  \bibnamefont {Craik}}\ and\ \bibinfo {author} {\bibfnamefont
  {S.}~\bibnamefont {Leibovich}},\ }\bibfield  {title} {\enquote {\bibinfo
  {title} {A rational model for {Langmuir} circulations},}\ }\href@noop {}
  {\bibfield  {journal} {\bibinfo  {journal} {Journal of Fluid Mechanics}\
  }\textbf {\bibinfo {volume} {73}},\ \bibinfo {pages} {401–426} (\bibinfo
  {year} {1976})}\BibitemShut {NoStop}%
\bibitem [{\citenamefont {Craik}(1982)}]{craik_1982}%
  \BibitemOpen
  \bibfield  {author} {\bibinfo {author} {\bibfnamefont {A.~D.~D.}\
  \bibnamefont {Craik}},\ }\bibfield  {title} {\enquote {\bibinfo {title}
  {Wave-induced longitudinal-vortex instability in shear flows},}\ }\href@noop
  {} {\bibfield  {journal} {\bibinfo  {journal} {Journal of Fluid Mechanics}\
  }\textbf {\bibinfo {volume} {125}},\ \bibinfo {pages} {37–52} (\bibinfo
  {year} {1982})}\BibitemShut {NoStop}%
\bibitem [{\citenamefont {Kawamura}(2000)}]{kawamura2000numerical}%
  \BibitemOpen
  \bibfield  {author} {\bibinfo {author} {\bibfnamefont {T.}~\bibnamefont
  {Kawamura}},\ }\bibfield  {title} {\enquote {\bibinfo {title} {Numerical
  investigation of turbulence near a sheared air--water interface. part 2:
  Interaction of turbulent shear flow with surface waves},}\ }\href@noop {}
  {\bibfield  {journal} {\bibinfo  {journal} {Journal of marine science and
  technology}\ }\textbf {\bibinfo {volume} {5}},\ \bibinfo {pages} {161--175}
  (\bibinfo {year} {2000})}\BibitemShut {NoStop}%
\bibitem [{\citenamefont {Veron}\ and\ \citenamefont
  {Melville}(2001)}]{veron2001experiments}%
  \BibitemOpen
  \bibfield  {author} {\bibinfo {author} {\bibfnamefont {F.}~\bibnamefont
  {Veron}}\ and\ \bibinfo {author} {\bibfnamefont {W.~K.}\ \bibnamefont
  {Melville}},\ }\bibfield  {title} {\enquote {\bibinfo {title} {Experiments on
  the stability and transition of wind-driven water surfaces},}\ }\href@noop {}
  {\bibfield  {journal} {\bibinfo  {journal} {Journal of Fluid Mechanics}\
  }\textbf {\bibinfo {volume} {446}},\ \bibinfo {pages} {25--65} (\bibinfo
  {year} {2001})}\BibitemShut {NoStop}%
\bibitem [{\citenamefont {Fujiwara}\ and\ \citenamefont
  {Yoshikawa}(2020)}]{fujiwara2020mutual}%
  \BibitemOpen
  \bibfield  {author} {\bibinfo {author} {\bibfnamefont {Y.}~\bibnamefont
  {Fujiwara}}\ and\ \bibinfo {author} {\bibfnamefont {Y.}~\bibnamefont
  {Yoshikawa}},\ }\bibfield  {title} {\enquote {\bibinfo {title} {Mutual
  interaction between surface waves and {Langmuir} circulations observed in
  wave-resolving numerical simulations},}\ }\href@noop {} {\bibfield  {journal}
  {\bibinfo  {journal} {Journal of Physical Oceanography}\ }\textbf {\bibinfo
  {volume} {50}},\ \bibinfo {pages} {2323--2339} (\bibinfo {year}
  {2020})}\BibitemShut {NoStop}%
\bibitem [{\citenamefont {Suzuki}(2019)}]{suzuki2019physical}%
  \BibitemOpen
  \bibfield  {author} {\bibinfo {author} {\bibfnamefont {N.}~\bibnamefont
  {Suzuki}},\ }\bibfield  {title} {\enquote {\bibinfo {title} {On the physical
  mechanisms of the two-way coupling between a surface wave field and a
  circulation consisting of a roll and streak},}\ }\href@noop {} {\bibfield
  {journal} {\bibinfo  {journal} {Journal of Fluid Mechanics}\ }\textbf
  {\bibinfo {volume} {881}},\ \bibinfo {pages} {906--950} (\bibinfo {year}
  {2019})}\BibitemShut {NoStop}%
\bibitem [{\citenamefont {Leibovich}(1977)}]{leibovich1977evolution}%
  \BibitemOpen
  \bibfield  {author} {\bibinfo {author} {\bibfnamefont {S.}~\bibnamefont
  {Leibovich}},\ }\bibfield  {title} {\enquote {\bibinfo {title} {On the
  evolution of the system of wind drift currents and {Langmuir} circulations in
  the ocean. {Part} 1. {Theory} and averaged current},}\ }\href@noop {}
  {\bibfield  {journal} {\bibinfo  {journal} {Journal of Fluid Mechanics}\
  }\textbf {\bibinfo {volume} {79}},\ \bibinfo {pages} {715--743} (\bibinfo
  {year} {1977})}\BibitemShut {NoStop}%
\bibitem [{\citenamefont {Couston}, \citenamefont {Jalali},\ and\ \citenamefont
  {Alam}(2017)}]{couston2017shore}%
  \BibitemOpen
  \bibfield  {author} {\bibinfo {author} {\bibfnamefont {L.-A.}\ \bibnamefont
  {Couston}}, \bibinfo {author} {\bibfnamefont {M.~A.}\ \bibnamefont
  {Jalali}},\ and\ \bibinfo {author} {\bibfnamefont {M.-R.}\ \bibnamefont
  {Alam}},\ }\bibfield  {title} {\enquote {\bibinfo {title} {Shore protection
  by oblique seabed bars},}\ }\href@noop {} {\bibfield  {journal} {\bibinfo
  {journal} {Journal of Fluid Mechanics}\ }\textbf {\bibinfo {volume} {815}},\
  \bibinfo {pages} {481--510} (\bibinfo {year} {2017})}\BibitemShut {NoStop}%
\bibitem [{\citenamefont {Lamb}(1932)}]{lamb1916hydrodynamics}%
  \BibitemOpen
  \bibfield  {author} {\bibinfo {author} {\bibfnamefont {H.}~\bibnamefont
  {Lamb}},\ }\href@noop {} {\emph {\bibinfo {title} {Hydrodynamics}}}\
  (\bibinfo  {publisher} {University Press},\ \bibinfo {year}
  {1932})\BibitemShut {NoStop}%
\bibitem [{\citenamefont {Holm}(1996)}]{holm1996ideal}%
  \BibitemOpen
  \bibfield  {author} {\bibinfo {author} {\bibfnamefont {D.~D.}\ \bibnamefont
  {Holm}},\ }\bibfield  {title} {\enquote {\bibinfo {title} {The ideal
  Craik-Leibovich equations},}\ }\href@noop {} {\bibfield  {journal} {\bibinfo
  {journal} {Physica D: Nonlinear Phenomena}\ }\textbf {\bibinfo {volume}
  {98}},\ \bibinfo {pages} {415--441} (\bibinfo {year} {1996})}\BibitemShut
  {NoStop}%
\bibitem [{\citenamefont {Stewart}\ and\ \citenamefont
  {Joy}(1974)}]{stewart1974hf}%
  \BibitemOpen
  \bibfield  {author} {\bibinfo {author} {\bibfnamefont {R.~H.}\ \bibnamefont
  {Stewart}}\ and\ \bibinfo {author} {\bibfnamefont {J.~W.}\ \bibnamefont
  {Joy}},\ }\bibfield  {title} {\enquote {\bibinfo {title} {{HF} radio
  measurements of surface currents},}\ }in\ \href@noop {} {\emph {\bibinfo
  {booktitle} {Deep sea research and oceanographic abstracts}}},\ Vol.~\bibinfo
  {volume} {21}\ (\bibinfo {organization} {Elsevier},\ \bibinfo {year} {1974})\
  pp.\ \bibinfo {pages} {1039--1049}\BibitemShut {NoStop}%
\bibitem [{\citenamefont {Longuet-Higgins}(1953)}]{longuet1953mass}%
  \BibitemOpen
  \bibfield  {author} {\bibinfo {author} {\bibfnamefont {M.~S.}\ \bibnamefont
  {Longuet-Higgins}},\ }\bibfield  {title} {\enquote {\bibinfo {title} {Mass
  transport in water waves},}\ }\href@noop {} {\bibfield  {journal} {\bibinfo
  {journal} {Philosophical Transactions of the Royal Society of London. Series
  A, Mathematical and Physical Sciences}\ }\textbf {\bibinfo {volume} {245}},\
  \bibinfo {pages} {535--581} (\bibinfo {year} {1953})}\BibitemShut {NoStop}%
\bibitem [{\citenamefont {Nicol{\'a}s}\ and\ \citenamefont
  {Vega}(2003)}]{nicolas2003three}%
  \BibitemOpen
  \bibfield  {author} {\bibinfo {author} {\bibfnamefont {J.~A.}\ \bibnamefont
  {Nicol{\'a}s}}\ and\ \bibinfo {author} {\bibfnamefont {J.~M.}\ \bibnamefont
  {Vega}},\ }\bibfield  {title} {\enquote {\bibinfo {title} {Three-dimensional
  streaming flows driven by oscillatory boundary layers},}\ }\href@noop {}
  {\bibfield  {journal} {\bibinfo  {journal} {Fluid Dynamics Research}\
  }\textbf {\bibinfo {volume} {32}},\ \bibinfo {pages} {119--139} (\bibinfo
  {year} {2003})}\BibitemShut {NoStop}%
\bibitem [{\citenamefont {Filatov}\ \emph {et~al.}(2016)\citenamefont
  {Filatov}, \citenamefont {Parfenyev}, \citenamefont {Vergeles}, \citenamefont
  {Brazhnikov}, \citenamefont {Levchenko},\ and\ \citenamefont
  {Lebedev}}]{filatov2016nonlinear}%
  \BibitemOpen
  \bibfield  {author} {\bibinfo {author} {\bibfnamefont {S.~V.}\ \bibnamefont
  {Filatov}}, \bibinfo {author} {\bibfnamefont {V.~M.}\ \bibnamefont
  {Parfenyev}}, \bibinfo {author} {\bibfnamefont {S.~S.}\ \bibnamefont
  {Vergeles}}, \bibinfo {author} {\bibfnamefont {M.~Y.}\ \bibnamefont
  {Brazhnikov}}, \bibinfo {author} {\bibfnamefont {A.~A.}\ \bibnamefont
  {Levchenko}},\ and\ \bibinfo {author} {\bibfnamefont {V.~V.}\ \bibnamefont
  {Lebedev}},\ }\bibfield  {title} {\enquote {\bibinfo {title} {Nonlinear
  generation of vorticity by surface waves},}\ }\href@noop {} {\bibfield
  {journal} {\bibinfo  {journal} {Physical Review Letters}\ }\textbf {\bibinfo
  {volume} {116}},\ \bibinfo {pages} {054501} (\bibinfo {year}
  {2016})}\BibitemShut {NoStop}%
\bibitem [{\citenamefont {Garrett}(1976)}]{garrett1976generation}%
  \BibitemOpen
  \bibfield  {author} {\bibinfo {author} {\bibfnamefont {C.}~\bibnamefont
  {Garrett}},\ }\bibfield  {title} {\enquote {\bibinfo {title} {Generation of
  {Langmuir} circulations by surface waves --- a feedback mechanism},}\
  }\href@noop {} {\bibfield  {journal} {\bibinfo  {journal} {Journal of Marine
  Research}\ }\textbf {\bibinfo {volume} {34}},\ \bibinfo {pages} {117}
  (\bibinfo {year} {1976})}\BibitemShut {NoStop}%
\bibitem [{\citenamefont {Abrashkin}\ and\ \citenamefont
  {Pelinovsky}(2022)}]{abrashkin2022gerstner}%
  \BibitemOpen
  \bibfield  {author} {\bibinfo {author} {\bibfnamefont {A.~A.}\ \bibnamefont
  {Abrashkin}}\ and\ \bibinfo {author} {\bibfnamefont {E.~N.}\ \bibnamefont
  {Pelinovsky}},\ }\bibfield  {title} {\enquote {\bibinfo {title} {Gerstner
  waves and their generalizations in hydrodynamics and geophysics},}\
  }\href@noop {} {\bibfield  {journal} {\bibinfo  {journal} {Physics--Uspekhi}\
  }\textbf {\bibinfo {volume} {65}},\ \bibinfo {pages} {453--467} (\bibinfo
  {year} {2022})}\BibitemShut {NoStop}%
\bibitem [{\citenamefont {Leibovich}(1983)}]{leibovich1983form}%
  \BibitemOpen
  \bibfield  {author} {\bibinfo {author} {\bibfnamefont {S.}~\bibnamefont
  {Leibovich}},\ }\bibfield  {title} {\enquote {\bibinfo {title} {The form and
  dynamics of {Langmuir} circulations},}\ }\href@noop {} {\bibfield  {journal}
  {\bibinfo  {journal} {Annual Review of Fluid Mechanics}\ }\textbf {\bibinfo
  {volume} {15}},\ \bibinfo {pages} {391--427} (\bibinfo {year}
  {1983})}\BibitemShut {NoStop}%
\bibitem [{\citenamefont {Craik}(1970)}]{craik1970wave}%
  \BibitemOpen
  \bibfield  {author} {\bibinfo {author} {\bibfnamefont {A.~D.~D.}\
  \bibnamefont {Craik}},\ }\bibfield  {title} {\enquote {\bibinfo {title} {A
  wave-interaction model for the generation of windrows},}\ }\href@noop {}
  {\bibfield  {journal} {\bibinfo  {journal} {Journal of Fluid Mechanics}\
  }\textbf {\bibinfo {volume} {41}},\ \bibinfo {pages} {801--821} (\bibinfo
  {year} {1970})}\BibitemShut {NoStop}%
\bibitem [{\citenamefont {Manna}\ and\ \citenamefont
  {Dhar}(2021)}]{manna2021modulational}%
  \BibitemOpen
  \bibfield  {author} {\bibinfo {author} {\bibfnamefont {S.}~\bibnamefont
  {Manna}}\ and\ \bibinfo {author} {\bibfnamefont {A.}~\bibnamefont {Dhar}},\
  }\bibfield  {title} {\enquote {\bibinfo {title} {Modulational instability of
  obliquely interacting capillary-gravity waves over infinite depth},}\
  }\href@noop {} {\bibfield  {journal} {\bibinfo  {journal} {Archives of
  Mechanics}\ }\textbf {\bibinfo {volume} {73}},\ \bibinfo {pages} {583--598}
  (\bibinfo {year} {2021})}\BibitemShut {NoStop}%
\bibitem [{\citenamefont {Yin}, \citenamefont {Pan},\ and\ \citenamefont
  {Chow}(2023)}]{yin2023modeling}%
  \BibitemOpen
  \bibfield  {author} {\bibinfo {author} {\bibfnamefont {H.}~\bibnamefont
  {Yin}}, \bibinfo {author} {\bibfnamefont {Q.}~\bibnamefont {Pan}},\ and\
  \bibinfo {author} {\bibfnamefont {K.}~\bibnamefont {Chow}},\ }\bibfield
  {title} {\enquote {\bibinfo {title} {Modeling “crossing sea state” wave
  patterns in layered and stratified fluids},}\ }\href@noop {} {\bibfield
  {journal} {\bibinfo  {journal} {Physical Review Fluids}\ }\textbf {\bibinfo
  {volume} {8}},\ \bibinfo {pages} {014802} (\bibinfo {year}
  {2023})}\BibitemShut {NoStop}%
\bibitem [{\citenamefont {Filatov}\ \emph {et~al.}(2018)\citenamefont
  {Filatov}, \citenamefont {Orlov}, \citenamefont {Brazhnikov},\ and\
  \citenamefont {Levchenko}}]{filatov2018experimental}%
  \BibitemOpen
  \bibfield  {author} {\bibinfo {author} {\bibfnamefont {S.~V.}\ \bibnamefont
  {Filatov}}, \bibinfo {author} {\bibfnamefont {A.~V.}\ \bibnamefont {Orlov}},
  \bibinfo {author} {\bibfnamefont {M.~Y.}\ \bibnamefont {Brazhnikov}},\ and\
  \bibinfo {author} {\bibfnamefont {A.~A.}\ \bibnamefont {Levchenko}},\
  }\bibfield  {title} {\enquote {\bibinfo {title} {Experimental simulation of
  the generation of a vortex flow on a water surface by a wave cascade},}\
  }\href@noop {} {\bibfield  {journal} {\bibinfo  {journal} {JETP Letters}\
  }\textbf {\bibinfo {volume} {108}},\ \bibinfo {pages} {519--526} (\bibinfo
  {year} {2018})}\BibitemShut {NoStop}%
\bibitem [{\citenamefont {Filatov}\ \emph {et~al.}(2022)\citenamefont
  {Filatov}, \citenamefont {Poplevin}, \citenamefont {Levchenko},\ and\
  \citenamefont {Parfenyev}}]{filatov2022generation}%
  \BibitemOpen
  \bibfield  {author} {\bibinfo {author} {\bibfnamefont {S.}~\bibnamefont
  {Filatov}}, \bibinfo {author} {\bibfnamefont {A.}~\bibnamefont {Poplevin}},
  \bibinfo {author} {\bibfnamefont {A.}~\bibnamefont {Levchenko}},\ and\
  \bibinfo {author} {\bibfnamefont {V.}~\bibnamefont {Parfenyev}},\ }\bibfield
  {title} {\enquote {\bibinfo {title} {Generation of stripe-like vortex flow by
  noncollinear waves on the water surface},}\ }\href@noop {} {\bibfield
  {journal} {\bibinfo  {journal} {Physica D: Nonlinear Phenomena}\ }\textbf
  {\bibinfo {volume} {434}} (\bibinfo {year} {2022})}\BibitemShut {NoStop}%
\bibitem [{\citenamefont {Parfenyev}\ \emph {et~al.}(2019)\citenamefont
  {Parfenyev}, \citenamefont {Filatov}, \citenamefont {Brazhnikov},
  \citenamefont {Vergeles},\ and\ \citenamefont
  {Levchenko}}]{parfenyev2019formation}%
  \BibitemOpen
  \bibfield  {author} {\bibinfo {author} {\bibfnamefont {V.}~\bibnamefont
  {Parfenyev}}, \bibinfo {author} {\bibfnamefont {S.}~\bibnamefont {Filatov}},
  \bibinfo {author} {\bibfnamefont {M.~Y.}\ \bibnamefont {Brazhnikov}},
  \bibinfo {author} {\bibfnamefont {S.}~\bibnamefont {Vergeles}},\ and\
  \bibinfo {author} {\bibfnamefont {A.}~\bibnamefont {Levchenko}},\ }\bibfield
  {title} {\enquote {\bibinfo {title} {Formation and decay of eddy currents
  generated by crossed surface waves},}\ }\href@noop {} {\bibfield  {journal}
  {\bibinfo  {journal} {Physical Review Fluids}\ }\textbf {\bibinfo {volume}
  {4}},\ \bibinfo {pages} {114701} (\bibinfo {year} {2019})}\BibitemShut
  {NoStop}%
\bibitem [{\citenamefont {Colombi}\ \emph {et~al.}(2022)\citenamefont
  {Colombi}, \citenamefont {Rohde}, \citenamefont {Schl{\"u}ter},\ and\
  \citenamefont {von Kameke}}]{colombi2022coexistence}%
  \BibitemOpen
  \bibfield  {author} {\bibinfo {author} {\bibfnamefont {R.}~\bibnamefont
  {Colombi}}, \bibinfo {author} {\bibfnamefont {N.}~\bibnamefont {Rohde}},
  \bibinfo {author} {\bibfnamefont {M.}~\bibnamefont {Schl{\"u}ter}},\ and\
  \bibinfo {author} {\bibfnamefont {A.}~\bibnamefont {von Kameke}},\ }\bibfield
   {title} {\enquote {\bibinfo {title} {Coexistence of inverse and direct
  energy cascades in {Faraday} waves},}\ }\href@noop {} {\bibfield  {journal}
  {\bibinfo  {journal} {Fluids}\ }\textbf {\bibinfo {volume} {7}},\ \bibinfo
  {pages} {148} (\bibinfo {year} {2022})}\BibitemShut {NoStop}%
\end{thebibliography}

%

\end{document}